\documentclass[sigconf]{acmart}
\usepackage[normalem]{ulem}  % TODO for Ben (08/27/25): REMOVE before CHI'26 Camera Ready!!!
\AtBeginDocument{%
  }

\copyrightyear{2026}
\acmYear{2026}
\setcopyright{cc}
\setcctype{by}
\acmConference[CHI '26]{Proceedings of the 2026 CHI Conference on Human Factors in Computing Systems}{April 13--17, 2026}{Barcelona, Spain}
\acmBooktitle{Proceedings of the 2026 CHI Conference on Human Factors in Computing Systems (CHI '26), April 13--17, 2026, Barcelona, Spain}
\acmDOI{10.1145/3772318.3791815}
\acmISBN{979-8-4007-2278-3/2026/04}

\acmSubmissionID{1522}

\newcommand{\revision}[1]{\textcolor{black}{#1}}
\newcommand{\ben}[1]{{\textcolor{black}{#1}}}

\newcommand{\quotes}[1]{``#1''}
\begin{document}

%%
%% The "title" command has an optional parameter,
%% allowing the author to define a "short title" to be used in page headers.
% TODO: Figure this out for camera ready
% {"originatingScript":"m2","payload":{"guid":"b50b3896-041c-405d-b12b-73757a39a649c70476","muid":"92d2bb17-92a6-42e3-b1f1-1498002580a062ed9b","sid":"e5486200-02da-4a43-889f-72a8baf5d7a0d96a9e"}}
% \title{A11yBits: An AI-Assisted Tangible Toolkit to Support DIY Solution Creation for People with Visual Impairments}
% \title{Co-Making with an AI Who's Also Blind: Challenges and Opportunities in Using AI for Co-Making Physical, DIY-AT for People with Visual Impairments}
\title[Not Seeing the Whole Picture]{Not Seeing the Whole Picture: Challenges and Opportunities in Using AI for Co-Making Physical DIY-AT for People with Visual Impairments}

%%
%% The "author" command and its associated commands are used to define
%% the authors and their affiliations.
%% Of note is the shared affiliation of the first two authors, and the
%% "authornote" and "authornotemark" commands
%% used to denote shared contribution to the research.
\orcid{0009-0000-4832-0458}
\author{Ben Kosa}
\email{bkosa@cs.wisc.edu}
\affiliation{%
  \institution{University of Wisconsin-Madison}
  \city{Madison}
  \state{Wisconsin}
  \country{USA}
}

\author{Hsuanling Lee}
% \authornote{Both authors contributed equally to this research.}
\email{hsuanling.lee@utdallas.edu}
\orcid{0009-0000-4832-0458}
\affiliation{%
  \institution{University of Texas at Dallas}
  \city{West Lafayette}
  \state{Indiana}
  \country{USA}
}

\author{Jasmine Li}
\email{jasmineli@utexas.edu}
\orcid{0009-0009-9208-6967}
\affiliation{%
  \institution{University of Texas at Austin}
  \city{Austin}
  \state{Texas}
  \country{USA}}

\author{Sanbrita Mondal}
\email{smondal4@wisc.edu}
\orcid{0000-0003-4454-8978}
\affiliation{%
  \institution{University of Wisconsin--Madison}
  \city{Madison}
  \state{Wisconsin}
  \country{USA}
}

\author{Yuhang Zhao}
\email{yuhang.zhao@cs.wisc.edu}
\orcid{0000-0003-3686-695X}
\affiliation{%
  \institution{University of Wisconsin--Madison}
  \city{Madison}
  \state{Wisconsin}
  \country{USA}
}

\author{Liang He}
\email{liang.he@utdallas.edu}
\orcid{0000-0003-4826-629X}
\affiliation{%
  \institution{University of Texas at Dallas}
  \city{Richardson}
  \state{TX}
  \country{USA}
}
%%
%% By default, the full list of authors will be used in the page
%% headers. Often, this list is too long, and will overlap
%% other information printed in the page headers. This command allows
%% the author to define a more concise list
%% of authors' names for this purpose.
\renewcommand{\shortauthors}{Kosa et al.}

%%
%% The abstract is a short summary of the work to be presented in the
%% article.
\begin{abstract}
Existing assistive technologies (AT) often adopt a one-size-fits-all approach, overlooking the diverse needs of people with visual impairments (PVI). Do-it-yourself AT (DIY-AT) toolkits offer one path toward customization, but most remain limited—targeting co-design with engineers or requiring programming expertise. Non-professionals with disabilities, including PVI, also face barriers such as inaccessible tools, lack of confidence, and insufficient technical knowledge. These gaps highlight the need for prototyping technologies that enable PVI to directly make their own AT. Building on emerging evidence that large language models (LLMs) can serve not only as visual aids but also as co-design partners, we present an exploratory study of how LLM-based AI can support PVI in the tangible DIY-AT co-making process. Our findings surface key challenges and design opportunities: the need for greater spatial and visual support, strategies for mitigating novel AI errors, and implications for designing more accessible AI-assisted prototypes.
\end{abstract}

\ccsdesc[500]{Human-centered computing~User interface toolkits}
\ccsdesc[500]{Human-centered computing~Accessibility systems and tools}

%%
%% Keywords. The author(s) should pick words that accurately describe
%% the work being presented. Separate the keywords with commas.
\keywords{Accessibility, Blind and low vision, DIY toolkit, Tangible modules }
%% A "teaser" image appears between the author and affiliation
%% information and the body of the document, and typically spans the
%% page.
\begin{teaserfigure}
  \includegraphics[width=\textwidth]{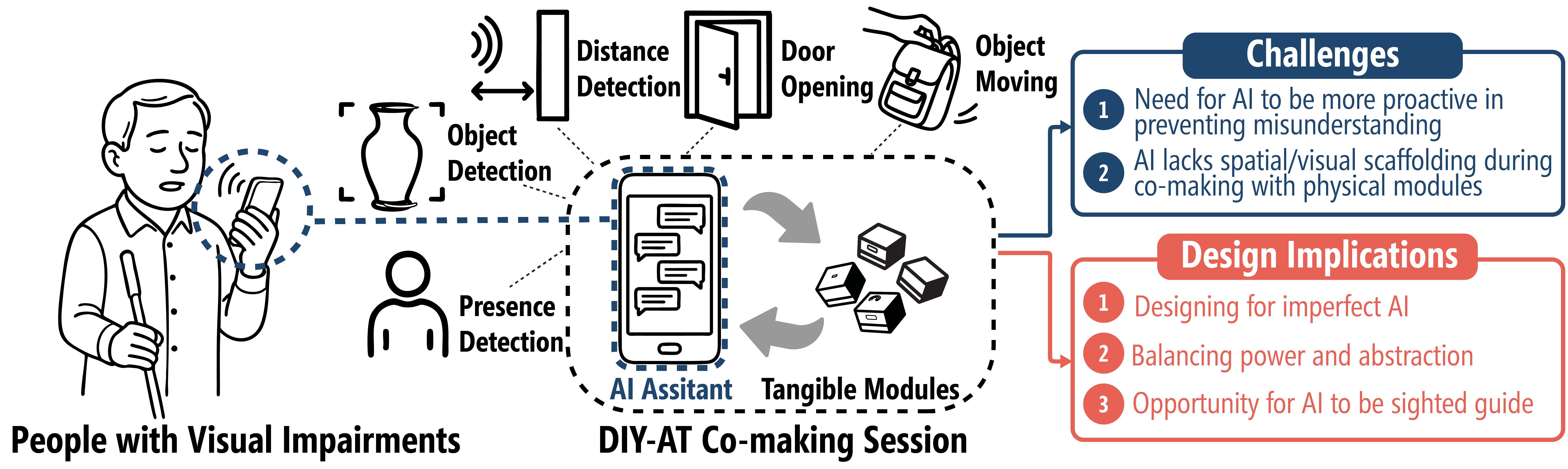}
  \caption{We examine how LLM-based AI supports PVI to create custom DIY-AT solutions and report the identified challenges of Co-Designing DIY-AT with AI for PVI and implications for designing AI-assist tools for PVI in creative tasks.}
  \label{fig:teaser}
\end{teaserfigure}

%%
%% This command processes the author and affiliation and title
%% information and builds the first part of the formatted document.
\maketitle

% Ben (11/06/2025): Hannah made some small changes to the intro that didn't make it to the first submission draft, so the first submission pdf and our draft now are slightly different.
% Ben (11/06/2025): the findings.tex was corrupted for some reason (first section repeated like 6 times), but I fixed it and verified that everything else is the same as our final draft that was submitted for the first round CHI'26 submission.
\section{Introduction}
For the estimated 285 million people worldwide who are affected by visual impairments \cite{steinmetz2021causes,WHO}, assistive technology (AT) represents a critical bridge to independence, productivity, and participation in society for them by supporting a wide range of daily challenges \cite{szpiro2016finding, jayant2011supporting, zhao2018face, zhao2018looks, williams2013pray, li2021non}. However, despite decades of technological advancement, the AT landscape remains constrained by a paradox: while individual needs are highly diverse and personal, AT solutions are predominantly standardized, expensive, and slow to evolve \cite{pape2002shaping,mennicken2014today,scherer2002change,verza2006interdisciplinary}. 

While accessibility research has increasingly used user-centered design \cite{abras2004user, laux1996designing, sanchez2008user}, traditional AT development follows a top-down, expert-driven model in which devices are designed by engineers and researchers, tested in controlled environments, and then distributed to users. 
This approach, while producing sophisticated devices, often fails to address the nuanced, context-specific, and evolving needs of individual users. 
%A screen reader that works perfectly in English may be inadequate for a multilingual user; a navigation aid optimized for urban settings may fail in rural environments; a reading device optimized for printed text may struggle with handwritten notes or diagrams. 
This mismatch results in high abandonment rates for AT, with poor fit, inflexibility, and high costs cited as primary factors \cite{hurst2013making,phillips1993predictors}. More critically, this model perpetuates dependency by positioning people with visual impairments (PVI) as passive recipients of technology rather than active creators and innovators.

This gap has inspired growing interest in alternative approaches such as do-it-yourself assistive technology (DIY-AT), which seeks to shift agency back to users themselves. 
DIY-AT empowers users to create solutions tailored to their own needs \cite{meissner2017yourself,hurst2013making}, often at lower cost and with greater accessibility than AT alternatives \cite{hook2014study,hurst2011empowering}. 
However, the creation process can itself be a significant barrier. Previous research has shown that non-professionals, especially those with disabilities, still face challenges in creating DIY-ATs, such as inaccessible maker spaces and tools \cite{meissner2017yourself}, lack of confidence \cite{hook2014study}, and insufficient technical knowledge \cite{herskovitz2023hacking}. 
%Even highly skilled Blind and Low Vision Software Professionals (BLVSPs) report what Braun et al.~\cite{braun2025} term the ``Double Hacker Dilemma''—possessing the expertise to address accessibility gaps but expending immense personal effort to do so, often duplicating work in isolation.
Existing DIY-AT toolkits for PVI have further been restricted to narrow use cases \cite{ducasse2016tangible,brown2012viztouch,minatani2019smart}, designed primarily for co-design activities rather than real-world implementation \cite{lefeuvre2016loaded} or dependent on advanced hardware/software skills \cite{blind_arduino_project}. 

Recent advances in large language models (LLMs) suggest new opportunities. AI assistants have a strong potential in addressing some of these key challenges in DIY-AT creation, namely allowing for more complex capabilities of DIY-AT through lowering the technological experience required \cite{herskovitz2023hacking} and helping empower creativity \cite{dang2025authoring, jeon2021fashionq, bennett2024painting}. Yet little is known about how PVI would actually engage with such assistants when co-making tangible, physical AT solutions.

To explore this, we developed an AI-assisted tangible DIY-AT toolkit, which comprises an \revision{LLM-based AI assistant}, \textit{A11yMaker AI}, and a set of tangible sensing and feedback modules for PVI to explore and assemble. 
% TODO FOR CAMERA READY: Mention that we based the tangible toolkit part of the system on our previous work, A11yBits.
The \revision{AI assistant} supports the entire DIY-AT creation through natural language: from brainstorming which modules to use, to configuring their behaviors, to managing complete programs. \revision{Using our AI-assisted toolkit as a study probe}, we conducted an exploratory study with nine PVI participants to examine: 
\begin{enumerate}
    \item How do PVI leverage an \revision{AI assistant} to brainstorm and co-make DIY-ATs using a distributed, tangible toolkit? \revision{What barriers do they face when co-making with AI?}
    \item What design considerations emerge when developing an \revision{LLM-based AI assistant} for PVI in the context of tangible, DIY-AT toolkits?  
\end{enumerate} 

Our exploratory study surfaces not only the opportunities of AI-assisted DIY-ATs but also the preliminary challenges PVI face when co-making with \revision{an LLM-based AI assistant} in physical embodied settings. 
\revision{Our findings reveal that PVI could create diverse DIY-ATs with the AI’s help, finding it useful as a patient tutor, toolkit expert, and technical abstraction. However, the assistant often failed to clarify toolkit limitations and resolve ambiguity—leading to partial or hallucinated solutions during physical co-making. The assistant also lacked the spatial and visual scaffolding participants needed to orient modules, verify the placement, or diagnose hardware failures during embodied co-making. These challenges highlight the importance of future LLM-based assistants in enabling the co-creation of tangible DIY-AT to be more proactive in surfacing physical limitations, making errors transparent, and evolving toward a multimodal, sighted guide.}

\noindent In summary, this paper contributes:

\begin{enumerate}
    \item \revision{The first exploration of integrating an LLM-based AI assistant in a tangible toolkit to support PVI in creating custom, tangible DIY-AT solutions.}
    \item Empirical insights from an exploratory study with nine PVI participants, \revision{revealing their usage, strategies, and challenges in using an LLM-based AI assistant} to brainstorm, configure, and assemble DIY-AT solutions, highlighting both opportunities and barriers in the co-making process.
    %highlights the usefulness of the toolkit in supporting the co-design of DIY-AT to meet a wide range of unique needs.
    \item Design insights for AI-assisted DIY-ATs, including the need for greater spatial and visual support, strategies for mitigating AI errors, and implications for more accessible AI-assisted prototyping.
\end{enumerate}

\section{Related Work}
% \ben{(08/31/2025) TODO Need to alter to match the new framing of this being an exploratory study and us presenting some preliminary challenges that PVI faced in co-making tangible, DIY-AT. Below is original UIST'25 version}
% \ben{(08/31/2025) Need to add relevant papers suggested by UIST reviews (plus Jaylin's error with AI paper)}
% In this section, we provide a brief background on technology abandonment for ATs to motivate DIY as a potential solution, prior work in creating DIY assistive technologies for PVI, and prior work in how GAI can help support PVI in creating and designing DIY-AT.
% \ben{(09/08/2025): Need to introduce VAT if I haven't already}

% % [Example:] In this section, we provide a brief background on sign languages and their linguistic features, describe existing approaches to sign language dictionaries, and outline current approaches to feature-based sign language search. 

% \ben{(08/22/2025): List of papers to add to related works / motivation}
% \begin{enumerate}
%     \item "EMBODIED INTELLIGENCE IN ASSISTIVE TECHNOLOGIES FOR THE VISUALLY IMPAIRED:ENHANCING INDEPENDENCE AND SOCIAL INCLUSION" -- Previous work that has pointed to the benefit of conversatioanl, EI (embodied intelligence) in AT for PVI.
%     \item TODO
% \end{enumerate}

\subsection{Technology Abandonment and Needs for DIY Solutions}

PVI face many challenges dealing with the physical environment and surrounding objects \cite{brady2013visual}. Various ATs have been designed to support PVI in daily tasks, such as navigation \cite{real2019navigation, ahmetovic2016navcog, sato2017navcog3, zhao2020effectiveness}, text reading \cite{bigham2010vizwiz, zhao2015foresee, ezaki2005improved, guo2016vizlens, Boldu2018FingerReader2}, and object recognition and localization \cite{Chen2022LiSee, huppert2021guidecopter, zhao2016cuesee, bigham2010vizwizlocate}. However, research has shown that more than 35\% of the ATs purchased by people with disabilities ended up being unused or abandoned \cite{kintsch2002framework, riemer1997factors}. 

Researchers have investigated the technology adoption and abandonment decision of people with disabilities and identified multiple factors that lead to AT abandonment, such as high cost \cite{phillips1993predictors, buehler2015sharing}, deep learning curve \cite{dawe2006desperately}, and perception of stigma caused by technology \cite{shinohara2011shadow}. Moreover, most ATs are designed based on the common needs of most users without sufficient customization options, thus failing to adequately meet a user's unique needs due to their varied abilities, living contexts, and personal experiences and preferences \cite{scherer2002change, verza2006interdisciplinary, copley2004barriers, pape2002shaping}. In addition, a user's needs may change over time due to changes in their health condition and life priorities, leading to the discontinuance of previously used technology \cite{phillips1993predictors, hurst2011empowering}. As such, Phillips and Zhao emphasized increasing user involvement in technology design and considering their long-term ability changes to reduce AT abandonment \cite{phillips1993predictors}. 

\revision{To address this problem, various ATs that allow for personalization and automatic adaptation have been developed to give users more flexibility to customize their experiences \cite{Sloan2010ThePO, wang2024gazePrompt, garrido2012personalized, stangl2021going, martin2018adaptiveWebNavigation, wen2024find}. However, they can only provide adaptation along a known dimension within a well-identified range, such as what eye movements should trigger word enlargement or speech in dynamic GUIs \cite{Sloan2010ThePO, wang2024gazePrompt} and what objects are important in scene description \cite{kacorri2017people, wen2024find}}. \revision{Users may still lack sufficient agency in curating a solution that completely fulfills their needs.} 

Compared to generic ATs, DIY-ATs allow non-professionals (\textit{e.g.,} people with disabilities, their family and friends, caregivers) to craft technologies that take full account of users' individual needs \cite{hurst2013making, hurst2011empowering, buehler2015sharing}, while avoiding the costly and complex process of professional need assessment during technology design \cite{hook2014study, hurst2013making}. \revision{It is also promising in addressing the long-tail problem of possibly diverse needs in adaptive ATs \cite{herskovitz2023hacking}}.

\subsection{Supporting DIY-AT Creation for People with Disabilities}

Maker and DIY communities strive to empower people from all backgrounds—not just those with formal training in engineering, computer science, or design—to participate in creating and designing technologies, guided by shared values of learning, democratization, and collaboration \cite{kuznetsov2010rise, taylor2016making}.
Hurst and Tobias were among the first to explore the idea of DIY-ATs, which emerged as a response to poor adoption rates stemming from limited discoverability, high costs, lack of suitable commercial options, evolving user needs, and the desire for greater customization \cite{hamidi2018participatory, hurst2011empowering, okerlund2019diy}. Early case studies of DIY-AT tend to highlight a co-design process, where domain experts helped develop solutions \cite{hurst2011empowering}. Research has since worked to make the prototyping process more accessible to all, including non-technical users. 
\revision{For example, there has been a rich body of work exploring consumer-grade fabrication tools (\textit{e.g.,} 3D printers and 3D modeling software) for DIY-AT making \cite{hofmann2014GripFab, kelly20153dPrintingProsthetic} and how to reduce technical barriers in using such tools \cite{stefanie2023style2Fab, mankoff2016clinicalMakerPerspectives, kane2015thingverse}.}
In this paper, we continue making the process of creating DIY-ATs more accessible and less technical, \revision{with a novel focus on leveraging the state-of-the-art LLM as co-making partners}. 

% Researchers have designed DIY toolkits and rapid prototyping technologies to support people with disabilities, care givers, and AT amateurs to engage in participatory creation and easily prototype DIY-ATs \cite{ambe2019older, moraiti2015empowering, hurst2013making, grierson2013noisebear}. Some toolkits were designed for rehabilitation therapists or care givers \cite{vandermaesen2014physicube, moraiti2015empowering, buehler2014coming, hamidi2019sensebox}. For example, Moraiti et al. \cite{moraiti2015empowering} explored the potential of the Skweezee system as a DIY toolkit for occupational therapists. The Skweezee system \cite{vanderloock2013skweezee} can convert everyday soft objects into smart objects and recognize squeeze-based gestures to support interactions. The researchers encapsulated the Skweezee system to a more novice friendly toolkit and provided a software interface that allowed the occupational therapists to define gestures and create tailored assistive solutions for their clients without the aid of a technical expert. Some DIY toolkits were designed to be used by people with disabilities \cite{ambe2019older, grierson2013noisebear, hamidi2018participatory}. For example, Ambe et al. designed Un-Kit \cite{ambe2019older}, a set of low-fidelity unfinished, unboxed electronic components that supported co-design experiences for older adults. With this toolkit, older adults can work with researchers in their familiar environment via in-home workshops to design and imagine AT solutions that fit their desire. 

More relevant to our project, researchers have designed prototyping toolkits specifically for PVI. 
Some focused on helping PVI create tactile artifacts, such as visualizations or maps \cite{ducasse2016tangible, brown2012viztouch, minatani2019smart}. For instance, Brown and Hurst developed \textit{VizTouch} \cite{brown2012viztouch}, a software system that enabled PVI to create tactile visualizations by converting a formula or an Excel sheet with data into an appropriate 3D printable physical visualization form.
Others have explored how to better support and engage PVI in co-designing assistive technologies with developers and engineers \cite{lefeuvre2016loaded, gadiraju2023offensive, glazko2024disabilityBias, giles2015imagining}. 
% For instance, Loaded Dice \cite{lefeuvre2016loaded} was a toolkit that facilitated participatory creation for PVI. It consisted of two wirelessly connected cubes---one embedded various sensors and the other embedded various actuators---which allowed PVI to match different input and output modalities during technology ideation and imagination in co-design activities. Giles and Linden \cite{giles2015imagining} also explored how involving an eTextile (i.e., electronic textiles) computing toolkit into a co-design workshop could help PVI share thoughts with each other and brainstorm future technologies.

However, prior research on PVI has mostly focused on co-design toolkits used in workshops, which require the engagement and guidance of researchers or experts. The tools only facilitated the brainstorming process but cannot generate ready-to-use solutions. While recent work has explored DIY-AT toolkits that allow PVI to create working AT themselves, these toolkits either require hardware and software development experience \cite{blind_arduino_project}, or are focused specifically on DIY-AT in software and on visual information filtering \cite{herskovitz2024diy}. \revision{The most relevant work is \ben{our previous project---}\textit{A11yBits} \cite{zhao2023a11yBits}---a tangible, distributed toolkit with a set of sensing and feedback modules that allow PVI to combine and assemble to create personalized DIY-ATs. While the ``plug-and-play'' mechanism of A11yBits can potentially reduce the technical barrier, PVI still need to spend significant effort in decision making (\textit{e.g.,} what modules to select to address a certain problem, how to integrate the solution into real-world environments). \ben{To remain accessible to non-technical users, A11yBits limits its module configuration (\textit{e.g.,} what temperature threshold a Temperature module should detect) to a small set of predefined behaviors exposed through simple buttons and dials, which constrains the range and flexibility of possible DIY-AT use cases.}}
% \ben{In making A11yBits simple enough to be used by non-technical users and non-visually programmable by simple buttons and dials, it came at the cost of limited functionality/domain of use cases.}

\revision{To lower the creativity and technical barriers \ben{while raising the ceiling of possible DIY-ATs}, Generative AI (GAI) presents a unique opportunity to co-design with non-technical PVI, abstract away technical complexity, and \ben{enable more flexible and expressive configuration of toolkit behaviors.} Our research seeks to seize this opportunity and investigate how to couple GAI with tangible toolkits to enable accessible, easy DIY-AT making for PVI.}  

%In contrast, we seek to design an accessible and easy-to-use toolkit that enables PVI to independently create DIY-ATs that can be used directly in daily life for a wide variety of potential use cases, without any assistance from technical experts.
%\ben{TODO: A bit long (~1000 characters). Shorten.}
%\revision{There has been some previous work that has started to explore creating easy-to-use toolkits for enabling PVI to create tangible, DIY-AT, such as A11yBits \cite{zhao2023a11yBits}. But in trying to abstract away technical complexity into a simple interface made up of knobs and switches, this approach limits the number of possible AT, which PVI participants highlighted as a limitation during user studies. Additionally, A11yBits demonstrated that PVI participants would like such a toolkit to have additional modules that can't easily be programmed or fully utilized through a simple button interface, such as a camera and IMU module that can support a wide range of computer vision and activity/movement detection. With the recent rise in exploring Generative AI as a way to abstract away technical complexity, it is a good possible solution to explore in allowing a tangible, DIY-AT to cover as wide of a range of various AT a PVI might to create without raising the technical complexity.}

% \input{fig_tab_alg/toolkit overview}
% \subsection{Generative AI as Assistive Technology}
\subsection{GAI in Supporting Creativity and DIY for PVI}
With the recent rise in LLMs, and GAI generally, there been a large number of work exploring the use of AI assistants, AI-assisted user interfaces, and human-AI collaboration in domains such as software development \cite{jetbrains2024ai}, education \cite{education}, domain-specific decision making \cite{zhang2018stockassistant, choudhury2023medisage, deldari2024auditnet}, making \cite{chen2023origami}, and general daily activities \cite{gao2024assistgpt}.
There has also been some recent work that has explored the potential for LLMs and GAI in creativity and ideation \cite{ivcevic2024artificial, o2024extending, chandrasekera2024can} in different domains, such as writing \cite{wan2024felt}, fashion \cite{jeon2021fashionq}, and DIY projects \cite{bercher2021yourself, behnke2019alice, li2024exploring}. Previous work surfaced general design considerations for LLM-powered conversational interfaces for sighted users, such as the importance of managing cognitive load, structuring turn-taking, and clearly communicating system uncertainty to support productive human–AI collaboration \cite{Amershi2019Guidelines}.

\begin{figure*}[h!]
  \includegraphics[width=\textwidth]{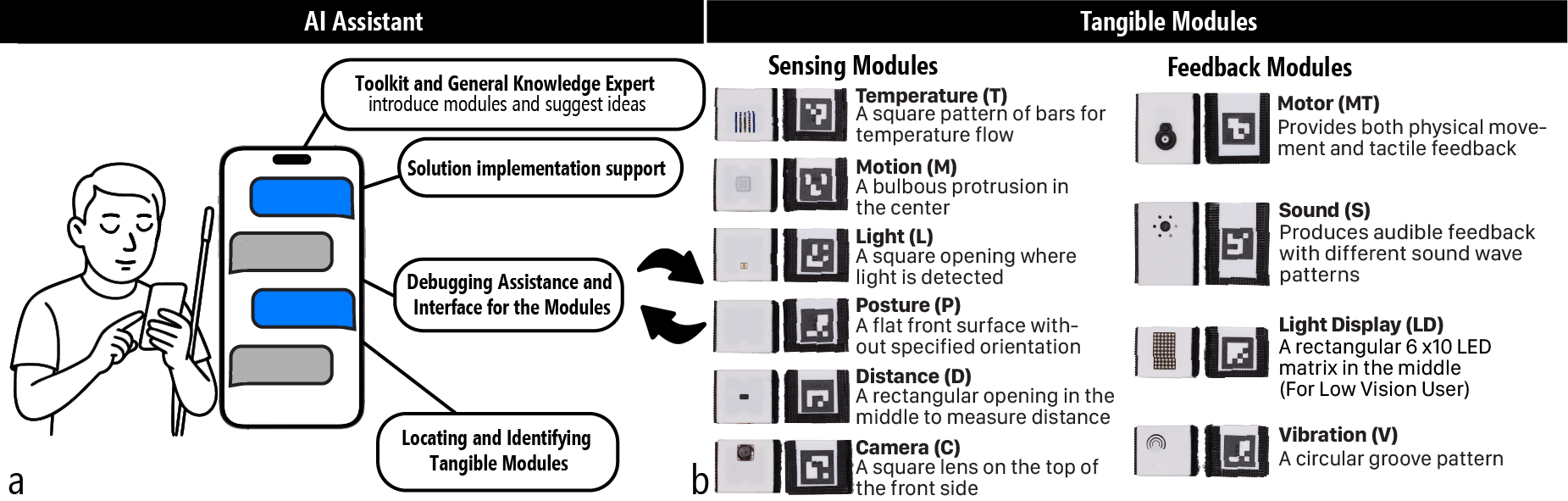}
  % \vspace{-1em}
  \caption{The DIY-AT toolkit consists of (a) an AI Assistant and (b) tangible modules.}
  \Description{A labeled diagram of the tangible DIT-AT toolkit showing two components: the AI assistant and the tangible modules. On the left, a smartphone interface for the AI Assistant is shown, connected to a Blind user holding a phone. Four labeled roles of the AI assistant are illustrated: introducing modules and suggesting ideas, supporting solution implementation, offering debugging help, and helping locate modules. On the right, there are six Sensing Modules—Temperature, Motion, Light, Posture, Distance, and Camera—each with a short description and a distinct tactile pattern or shape. Next to them are four Feedback Modules—Motor, Sound, Light Display, and Vibration—with brief explanations of their output types. }
  % \vspace{-1em}
  \label{fig:modules}
  
\end{figure*}

Within the domain of PVI and accessibility, most research has only focused on the use of GAI in visual description \cite{zhang25a11yshape,huh24designchecker}. PVI face challenges in perceiving their surroundings due to the absence of visual cues, and so naturally,
recent advances in multi-modal large language models (MLLMs) have spurred a large body of work in exploring how AI can help visually interpret the visual world for PVI \cite{ahmetovic2020recog, gonzalez2024investigating, hong2022blind, morrison2023understanding, mukhiddinov2022automatic, penuela2025towards}, and even commercial tools like \textit{Seeing AI} \cite{SeeingAI} and \textit{Be my AI} \cite{BeMyAI}.

Recently, research has started to explore how conversational AI can benefit PVI outside of just visual interpretation \cite{yang2024viassist, zhao2024vialm, glazko2023autoethnographic, adnin2024look}, including a growing body of work over the past few years exploring GAI supporting PVI creativity in visual content creation
\cite{huh2023genassist, lee2024altcanvas, huh24designchecker,zhang25a11yshape} and art \cite{chheda2025artinsight, bennett2024painting} as well as enhancing productivity applications for PVI \cite{minoli2024blvaiproductivity}. \revision{Prior work has identified several design considerations and issues for GAI-based interfaces for PVI, including the need for multimodal input (\textit{e.g.,} voice) and screen-reader–accessible interactions \cite{adnin2024look}, the risk of biased or ableist language in commercial LLMs due to non-representative training data and the lack of additional alignment \cite{adnin2024look, glazko2023autoethnographic}, and the difficulty PVI face when challenging visual hallucinations without access to visual cues \cite{misfittingwithAI}. Moreover, Aflatoony \textit{et al.} \cite{aflatoony2025aifabrication} have also examined how GAI can assist non-expert makers (\textit{e.g.,} clinicians) in DIY-AT fabrication, while highlighting challenges such as AI-generated generic designs and limited disability-specific training data.
However, prior research has focused on AI-assisted software-based creativity for PVI or fabrication workflows for sighted caregivers. No work has examined how GAI can scaffold the embodied process of physical DIY-AT making with distributed, tangible toolkits, or how these interfaces must adapt when supporting PVI who cannot visually validate intermediate steps.}

\section{AI-assisted Tangible DIY-AT Toolkit}

\revision{Our research seeks to explore how GAI interfaces can support PVI in DIY-AT creation with tangible toolkits, revealing barriers and implications to inspire future AI-assisted DIY-AT co-making.} To provide PVI hands-on experiences, we developed an AI-assisted tangible DIY-AT toolkit as our study probe. The system comprises an \revision{LLM-based AI assistant}---A11yMaker AI---as well as a set of tangible sensing and feedback modules \revision{for PVI to select, combine, and configure to create DIY-AT solutions} (Figure~\ref{fig:modules}). \revision{Unlike prior work that requires PVI to design and assemble solutions by themselves (\textit{e.g.,} A11yBits \cite{zhao2023a11yBits})}, A11yMaker AI enables PVI to co-design and program DIY-AT solutions through natural language—from brainstorming and selecting modules to configuring and managing completed programs.
% \ben{QUESTION FOR LH and YZ: Does it make sense to add back Figure 2 for further explaining AI implementation details?}

\subsection{Tangible Modules}
We first developed a set of sensing and feedback tangible modules to represent state-of-the-art tangible toolkits. Inspired by \revision{commercial} electronic kits (\textit{e.g.}, Arduino Starter Kit, Snap Circuits\textsuperscript{\textregistered}) and \revision{prior research on tangible DIY-AT toolkits for PVI \cite{zhao2023a11yBits, lefeuvre2016loaded, blind_arduino_project}}, % and existing AT (\textit{e.g.}, Seeing AI\footnote{\url{https://www.seeingai.com/}} \footnote{\url{https://www.bemyeyes.com/be-my-ai}}), 
we developed six sensing modules (\textit{Camera}, \textit{Distance}, \textit{Motion}, \textit{Light}, \textit{Posture}, \textit{Temperature}) and four feedback modules (\textit{Vibration}, \textit{Sound}, \textit{Light Display}, \textit{Motor}) (Figure~\ref{fig:modules}) \revision{to act as our tangible DIY-AT toolkit probe.} 
\revision{We select these multimodal modules to meet PVI's diverse sensory needs and preferences across different environments based on prior literature (\textit{e.g.,} using Camera Module to detect humans and objects \cite{bigham2010vizwiz, HamiltonFletcher2019SoundSightAMA, Gamage2023WhatDBA})} and cover a wide range of input and output capabilities to open up a broad creativity space for PVI. 
%to align with prior DIY-AT toolkits for PVI \cite{zhao2023a11yBits, lefeuvre2016loaded} and support the BLV community’s varied feedback preferences across different environments and levels of visual ability.}
% \revision{We chose to include our six sensing modules based on sensors and capabilities that have been included by previous DIY-AT toolkits designed for PVI \cite{zhao2023a11yBits, lefeuvre2016loaded}. Similarly, we include sound, haptic, LED, and actuated feedback in our toolkit based on previous DIY-AT toolkits for PVI \cite{zhao2023a11yBits, lefeuvre2016loaded}, based on the fact that different people in the BLV community have different preferences in feedback modality depending on the environment and their visual ability.}
These modules function as accessible, reconfigurable building blocks that PVI can mix and match to rapidly prototype assistive devices, similar to the DIY toolkit mechanism in prior work \cite{zhao2023a11yBits}. The functionality of each module is specified in Appendix~\ref{appendix:tangible}.

All modules were built with commodity electronics.
% TODO FOR CAMERA READY: In considering the modules to use for our probe, we followed a similar design to A11yBits, taking into consideration some of the additional modules and improvements recommended by PVI particpants (e.g. this part in the paper ("Meanwhile, participants pointed out the drawbacks of A11yBitsand suggested valuable improvements. First, more sensing and feed-back modules were suggested to support a broader range of complexand diverse daily tasks, such as a Camera module for text and objectrecognition, an Inertial Measurement Unit (IMU) module to track the user’s behaviors")
Each module has a 3D-printed case with three layers (Figure~\ref{fig:exploded} in Appendix), including an  
\textbf{Electronic Cover} housing the sensor or actuator, shaped with tactile cues for recognition;  
a \textbf{Circuit-control Body} with an ESP32-S3 microcontroller, a custom PCB, battery charging, a piezo buzzer that generates audio feedback for PVI to locate the module, and Velcro strips for attachment; and a
\textbf{Power Base} with a rechargeable battery, Braille labels, Velcro strips for orientation and attachment, and a central ArUco marker for camera-based identification.  
Layers slot together for a compact, durable enclosure. Additional implementation details are in Appendix~\ref{appendix:implement}.

\revision{Note that we developed these modules to be an example of tangible DIY-AT toolkits, serving as the technology context in our study to explore how PVI co-make with an AI assistant. While the toolkit can be expanded by adding more modules and functionalities, our findings on AI opportunities and barriers as co-making partner should still apply.}

\begin{figure*}[h]
  \includegraphics[width=0.8\linewidth]{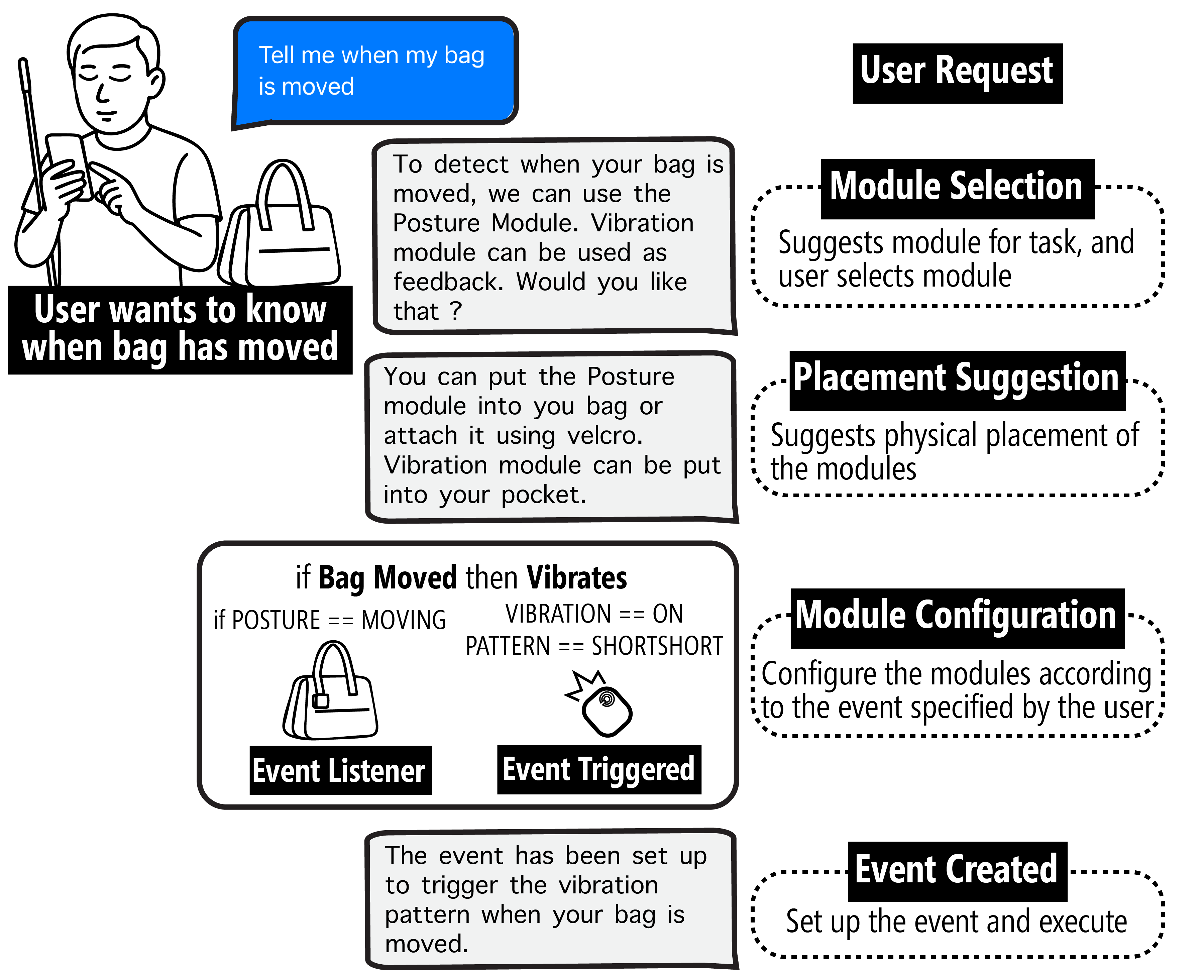}
  % \vspace{-0.5em}
  \caption{An step-by-step example of the participant using the A11yBits toolkit to know when their bag gets moved.}
  \Description{A step-by-step visual showing how a participant uses the A11yBits toolkit to detect if their bag is moved. On the left, a Blind user requests, “Tell me when my bag is moved.” The AI assistant responds by recommending the Posture Module to detect motion and the Vibration Module for feedback. It suggests placing the Posture Module inside the bag and the Vibration Module in the user’s pocket. A logic rule is created: “If POSTURE == MOVING, then VIBRATION == ON with SHORTSHORT pattern.” The event is configured and activated, completing the setup. Labeled steps on the right describe each phase: User Request, Module Selection, Placement Suggestion, Module Configuration, and Event Created.}
  % \vspace{-1em}
  \label{fig:scenario}
\end{figure*}

\subsection{A11yMaker AI}
% \ben{I'm realizing that the only really important part of adding more details about the A11yMaker AI is clarifying that the AI had a reasonable amount of context in the system prompt about how the modules worked and their capabilities, to help clarify that we gave it a reasonable amount of context about system limitations.}
\revision{To enable easy and accessible AI-assisted co-making}, we integrated a general-knowledge, LLM-based assistant into a smartphone app, allowing PVI to interact via speech. 
\revision{Inspired by prior expert-PVI co-making work \cite{zhao2023a11yBits, lefeuvre2016loaded}, we designed A11yMaker AI to: (1) be a toolkit and knowledge expert to support brainstorming; (2) be able to implement solutions with the toolkit and abstract away technical/programming details for non-technical PVI users; (3) make module details available to PVI users when they want more control or need to debug; and (4) be aware of the user's visual impairments and provide spatial, visual information of the modules (\textit{e.g.,} location and orientation).} For our prototype, we used GPT-4o as the model and designed the AI assistant based on the four aspects below:

\paragraph{Brainstorming partner and toolkit expert}
\revision{To help make tangible DIY-AT toolkits more accessible for non-technical PVI users, we designed our A11yMaker AI assistant to be both an always-available toolkit expert and a brainstorming partner, introducing modules to new users, explaining functions, and suggesting ideas to match user needs.
We implemented this by creating a specialized system prompt that included instructions for how to be a good brainstorming partner (\textit{e.g.,} using both divergent and convergent thinking principles \cite{liu2003towards, ciolfi2016articulating}% \yuhang{need explanation and also quote some of your actual prompt; i think you want to thorough about how you design the prompts, especially as brainstorming partners}
). We also provided the AI assistant with sufficient knowledge of the toolkit. Details about the capabilities of each module are included separately in the descriptions of the functions that are available to A11yMaker AI, which are automatically integrated into the LLM's context. We include the main system prompt in Appendix~\ref{appendix:system-prompt}}.
% \ben{Question for YZ and LH: Does this make sense? We can also include all of these functions and their descriptions in the appendix, but I think we'd have to put them into a table. I don't think the reviewers want that much detail?}
% \ben{, not including the additional function call descriptions that describe the capabilities of each module,}

% Design choices / impelementation details:
%   1. Open AI's "Tool Call" system, which describes the toolkit functionality, the functionality of individual module, and the functions that are available to A11yMaker AI to call to program and interact with different modules in the toolkit. This is all part of the system prompt, but we don't include it in the Appendix.
\paragraph{Solution implementation}
% - "Only mention one step at a time"
\revision{The A11yMaker AI assistant abstracts programming details by accessing real-time sensing data from sensing modules and configuring event-triggered programs that drive user-specified feedback modules. The AI assistant interfaces with the toolkit's backend to determine which pre-programmed functions to call and prompts the user for the parameters needed to configure their solution, thereby hiding low-level module programming. 
This allows users to create programs through natural language that trigger feedback (vibration, sound, or light) based on ``events'' defined by a set of conditionals (Figure \ref{fig:scenario}).}
%\revision{Our A11yMaker AI assistant prototype specifically makes use of OpenAI's Tool calling system \footnote{https://platform.openai.com/docs/guides/function-calling}}
\revision{For example, to receive a vibration when someone moves their bag, the event is defined as the bag moving (POSTURE$\_$MODULE == MOVING), which triggers a specified vibration pattern with the Vibration Module (VIBRATION == ON \& PATTERN == SHORT\_SHORT). More complex events can combine multiple conditionals, such as generating an audio alert when the camera detects a person within one meter who is about to collide with the user (CAMERA DETECTS PERSON \& DISTANCE $\leq$ 1m) (see Figure \ref{fig:scenario} for an example). To enable more intelligent capabilities, the A11yMaker AI assistant also supports AI-powered recognition functions based on the Camera Module (\textit{e.g.,} face recognition, object detection) by using OpenAI's Tool calling system\footnote{OpenAI Function Calling: https://platform.openai.com/docs/guides/function-calling}}.

% Design choices / impelementation details:
%   1. 
\paragraph{Configuring and debugging the module interface}
% Through wireless communication with each module, the AI reported states (e.g., battery level, sensor values) and relayed commands (e.g., trigger feedback, query distance). This let users quickly probe parameters, such as typical light near a candle when designing reminders.
\revision{To enable solution control and debugging, A11yMaker AI verbally communicates the module status and information upon PVI's request. The user can ask the AI assistant for module-specific information, such as battery level, connection status, and current sensor readings. The AI assistant connects to each module wirelessly and accesses module states as needed to respond.
The user can also interact with individual modules through the AI assistant for testing and control purposes. For example, requesting the current readings from a sensing module (\textit{e.g.,} \quotes{what is the current temperature of the room}) or triggering a certain feedback pattern with a feedback module. 
This allows users to quickly probe parameters, for example, testing out the typical light near a candle with the Light Module when designing reminders for putting out a candle.}
% This capability is particularly useful during co-design, enabling the user to quickly probe relevant parameters for their programs---such as typical light intensity near a candle when designing reminders when the candles are left on (\ref{use_cases-distributed}), or evaluating preferred feedback types and patterns.}

% \paragraph{\textbf{Locating and identifying modules}} % Pre-revision
\paragraph{Support Visual/Spatial Awareness for PVI}
% - "Always make sure to offer to use the "beepModule" tool to help them locate each module
% - Including instructions to be mindful that PVI users may not be initially aware of the fact that some of the modules are direction specific (i.e. the sensor only points in one direction) and to include this guidence when setting up DIY-AT with the toolkit.
%      - "Consider hardware limitations (e.g. sensor placement, orientation)."
%      - "Help users think through practical mounting and positioning options:** Ask where they plan to place the module and ensure it makes sense for their task."
\revision{Because making tangible DIY-ATs with a toolkit involves spatial constraints—for example, modules may need specific orientations or be distributed across a workspace—we designed A11yMaker AI to help PVI identify and locate modules.}
For PVI who are unfamiliar with Braille, the AI assistant could scan the ArUco marker on each module via the smartphone's built-in camera to receive verbal descriptions of the module. A module could also emit short sounds to help users locate it when they make a request to the assistant.
\revision{Importantly, we encoded instructions in the system prompt to ensure that A11yMaker AI discusses critical and highly visual details, such as module placement and orientation, with PVI during co-making.}

\section{Study Method}
%\rough{[Maybe alt: "Evaluation Protocol]}
Using our tangible DIY-AT toolkit as a probe, we explore how a conversational AI assistant can support PVI in co-making tangible DIY-ATs. We conducted a lab-based think-aloud study with nine PVI. 

\subsection{Participants}
We recruited nine participants with visual impairments through local disability networks and online communities. The average age was 49.6 years ($SD=20.9$); five
%55.6\% 
identified as female and four
%44.4\%
as male. The participants varied in their previous experience with assistive technologies and voice assistants (Table \ref{tab:participantnew}). \revision{To be considered eligible for our study, participants had to be over 18 years old and have at least 20/400 visual acuity.}

\begin{table*}[t]
\caption{Individual study participant information, including gender, age, visual ability, and self-reported experience with assistive technology (AT) and AI.}
\label{tab:participantnew}
\centering
\normalsize
\begin{tabular}{p{0.05\textwidth} p{0.10\textwidth} p{0.20\textwidth} p{0.30\textwidth} p{0.25\textwidth}}
\toprule
\textbf{ID} & \textbf{Age / Sex} & \textbf{Visual Ability} & \textbf{Experience with AT} & \textbf{Experience with AI} \\
\midrule
P1 & 26 / F & Low vision & Cane, screen reader & Seeing AI \\
P2 & 71 / M & Blind & Screen reader & Bixby Vision, Gemini \\
P3 & 82 / M & Low vision & Cane, magnifier & None \\
P4 & 55 / M & Low vision & Screen reader, text-to-speech tools & Seeing AI \\
P5 & 32 / M & Low vision & Cane, magnifier, color adjustment, bump dots & Bixby Vision, Seeing AI, Gemini, Copilot \\
P6 & 44 / F & Blind & Cane, screen reader & None \\
P7 & 30 / F & Low vision & Screen reader, VoiceOver, braille display & Seeing AI, Be My Eyes, ChatGPT \\
P8 & 71 / F & Blind & Magnifier, CCTV & None \\
P9 & 35 / F & Low vision & Cane, screen reader, VoiceOver, braille display, Alexa, GPS & Seeing AI, Be My Eyes, Meta AI \\
\bottomrule
\end{tabular}
\end{table*}

% \begin{table}[H]
% \caption{Individual study participant information, including their gender, age, visual ability, self-reported experience with assistive technology and AI.}
% \footnotesize
% \centering
% %\caption{Participant demographics}
% %\vspace{-1.5em}
% \begin{tabular}{p{1.2em}p{3em}p{6.5em}p{6em}p{6em}}
% \toprule
% \textbf{ID} & \textbf{Age/ Sex} & \textbf{Visual Ability} & \textbf{Experience with AT} & \textbf{Experience with AI} \\
% \toprule
% P1  & 26/F & Low vision & Cane, Screen reader & Seeing AI \\
% P2  & 71/M & Blind & Screen reader & Bixby Vision, Gemini \\
% P3  & 82/M & Low vision & Cane, Magnifier & No experience\\
% P4  & 55/M & Low vision & Screen reader, Text-to-Speech Tools & Seeing AI \\
% P5  & 32/M & Low vision & CAne, Magnifier, Color adjustment, Bump dots & Bixby Vision, Seeing AI, Gemini, Copilot\\
% P6  & 44/F & Blind      & Cane, Screen reader & No experience\\
% P7  & 30/F & Low vision & Screen reader, Voice over, Braille display & Seeing AI, Be My Eyes, ChatGPT \\
% P8  & 71/F & Blind  & Magnifier, CCTV & No experience\\
% P9  & 35/F & Low vision & Cane, Screen reader, Voice over, Braille display, Alexa, GPS & Seeing AI, Be My Eyes, Meta AI \\
% \bottomrule
% \end{tabular}
% \vspace{0.5em}
% \label{tab:participantnew}
% \vspace{-2.0em} % Reduce space after the whole float
% \end{table}

\subsection{Procedure} 
Each session lasted about 150 minutes and followed four phases: a pre-study interview, a toolkit tutorial with a sample task, two solution-building tasks, and a post-task interview. The interview script is in Appendix~\ref{appendix:post-interview}. %\yuhang{attach interview script to appendix and refer to them here.}

\paragraph{Pre-Study Interview} We conducted a semi-structured interview, asking about participants' demographic details, information about their visual conditions, and prior experiences with ATs and AI-powered tools to contextualize their use of A11yMaker AI.

\paragraph{Toolkit Tutorial and Sample Task} Participants were introduced to the toolkit through a guided tutorial covering the modules and how to interact with the AI assistant. To familiarize them with the full flow, we walked through a predefined example task (\textit{e.g.,} receiving feedback when a bag is moved), keeping researcher guidance minimal to encourage independent reasoning.

\paragraph{Solution-Building Tasks} Participants identified personal accessibility challenges and used the toolkit to prototype two solutions. They were asked to think aloud, while researchers observed key interactions, strategies, and moments of confusion or insight and took notes. During the task, participants were not allowed to ask researchers questions about the toolkit, thus only relying on the AI assistant.
% During the post-task interview, researchers asked follow-up questions about participants' challenges, strategies, and underlying rationales.  

\paragraph{Post-Task Interview} \revision{Right after completing the solution-building tasks, we conducted a semi-structured interview where participants reflected} on the usability, effectiveness, real-world feasibility and limitations \revision{of the AI assistant}, and provided feedback and suggestions.

\subsection{Analysis}
Sessions were recorded (with consent) and transcribed via Zoom, and transcripts were corrected for accuracy. We also captured smartphone screen recordings and extracted all participant–AI conversations. Across nine sessions, we collected 486 quotes about participants’ experiences and interactions. \revision{We also used video recordings, along with AI-participant conversation logs and system logs from each session, to generate notes and descriptions for non-verbal interactions that didn't appear on the transcript, as well as system and AI errors that affected participants' co-making outcomes and experiences. We inserted these observational notes into the transcripts with timestamps.}

Using reflexive thematic analysis \cite{saldana2021coding, braun2006using}, two researchers independently coded transcripts from two participants to develop an initial codebook; one researcher then coded the remainder, with iterative updates by consensus. Codes were grouped into themes and sub-themes aligned with our research focus, including groupings based on the different observed \revision{use cases and assembled solutions generated with the AI-assisted tangible toolkit as well as usage strategies, concerns, and improvement suggestions for the AI assistant}.

\section{Findings}

During the study, participants co-made 14 DIY-ATs with A11yMaker AI, spanning navigation, item search, and environmental awareness tasks (Table \ref{tab:participant_scenarios}). We found that the \revision{AI assistant played critical roles} as a tutor, toolkit expert, and brainstorming partner throughout the physical DIY-AT co-making process, enabling participants to navigate the complexity of the distributed toolkit. However, our study also revealed significant challenges in PVI-AI co-making: we observed A11yMaker AI often failed to clarify the toolkit limitations, resolve ambiguity, or avoid hallucinating capabilities---issues that point to design opportunities for more proactive, transparent, and accessible AI support in tangible co-design. In this section, we \revision{present the benefits and} challenges of A11y Maker AI to surface key design considerations for future GAI co-making agents for PVI.

\renewcommand{\arraystretch}{1.5}
\begin{table*}[t]
\caption{Detailed use scenarios, proposed solutions, and modules identified and utilized by participants during the user study.}
\small
\centering
\begin{tabular}{p{2em}p{15em}p{20em}p{8em}p{8em}}
\toprule
\textbf{ID} & \textbf{Scenario} & \textbf{Solution Description} & \textbf{Sensing Module Used} & \textbf{Feedback Module Used} \\
\toprule
P1 & Navigating to the bus stop & User will receive a vibration alert when the camera sees a sign (general signs). & Camera & Vibration \\
\cline{2-5}
   & Determining if candles have been blown out & Using a light sensor to monitor changes in light intensity. & Light & LED Display \\
\midrule
P2 & Creating a tool belt that tells distance from large objects & Distance module attached to user via Velcro belt; sound alerts within 2 meters. & Distance & Sound \\
\midrule
P3 & Finding a TV in a store & Vibration alert triggered when the camera detects a TV. & Camera & Vibration \\
\midrule
P4 & Detecting a person at the door & Sound alert triggered when the camera detects a person at the door. & Camera & Sound \\
\cline{2-5}
   & Identifying a specific bus & OCR used to scan bus text for identification (unfinished task). & Camera & N/A \\
\midrule
P5 & Notifying the user when a person appears & Sound alert triggered when the camera detects a person. & Camera & Sound \\
\cline{2-5}
   & Reminding the user if lights are on/off & Sound alert when detected light intensity crosses threshold. & Light & Sound \\
\midrule
P6 & Finding shampoo in a grocery store & Vibration alert triggered when the camera detects a shampoo bottle. & Camera & Vibration \\
\midrule
P7 & Finding a dropped water bottle & Sound alert triggered upon camera detection of water bottle. & Camera & Sound \\
\midrule
P8 & Detecting nearby obstacles & Vibration alert triggered when an obstacle is detected within 1000 mm. & Distance & Vibration \\
\midrule
P9 & Finding an electrical outlet & Vibration alert triggered when the camera detects an outlet. & Camera & Vibration \\
\cline{2-5}
   & Finding a cell phone & Camera detects cell phone, triggers vibration; distance module measures proximity & Camera & Vibration \\
\cline{2-5}
   & Monitor room temperature and provide periodic updates & Temperature module tracks temperature; periodic sound alerts. & Temperature & Sound \\
\bottomrule
\end{tabular}
% % \vspace{0.5em}
\label{tab:participant_scenarios}
% \vspace{-2.0em}
\end{table*}

\revision{\subsection{Roles of AI assistant in Co-making}}
\revision{Through observing participants co-make physical, custom AT solutions with the AI assistant, we found that the AI assistant played several useful roles: acting as a patient tutor and toolkit expert, and helping participants translate needs into workable DIY-AT configurations.}

%\revision{\paragraph{A11yMaker AI as a useful, always-available teacher and toolkit expert in the tangible DIY-AT process}}
% OLD
% Participants often treated the assistant as a just-in-time trainer: asking how modules worked, what a configuration would do, or what to try next (P1, P2, P4, P5, P9). This was particularly valued when scheduling with human trainers was difficult or participants preferred not to impose on others. As P2 put it, \textit{``My resistance would be to use a live person, as I don't want to take up other people's time… I try and do stuff myself''} (P2). When P5 raised a future scenario (home dialysis), the AI listed potential risks, then clarified: \textit{``A11yBits is not designed for medical treatment, but I can help brainstorm ways to enhance accessibility or monitor conditions around your setup''} (A11yMaker AI). These kinds of exchanges show how participants used the assistant as an always-available tutor that answered questions, clarified boundaries, and kept exploration moving.

% P1 asking the AI to remind them how the LED module worked
\subsubsection{AI as always-available toolkit expert and trainer.} \revision{Participants often treated the AI assistant as a just-in-time trainer and toolkit expert (P1, P2, P4, P5, P9): asking how modules worked when brainstorming a DIY-AT (\textit{e.g.,} P9 asked the AI about what objects the Camera Module could detect, and P5 asked how many events would the toolkit allow them to chain together) or requesting clarification on how their built DIY-AT worked (P1, P2, P4). P2 highlighted the benefits of the AI assistant being always-available: \textit{``it was nice to be able to ask the AI questions when I had them...when learning new AT, like my JAWS screen reader, I found it really important to have hands-on help. However, it’s only like availability based. And so my availability and [the expert's] and to make them line up is not that easy and so that would be my biggest thing is just like these are tools that I need and yet it’s really hard to find the time to learn how to use them properly...It's nice that you don't need to schedule time with the AI.''} Some participants (P1, P9) also felt an always-available AI expert made them more independent, \textit{``My resistance would be to use a live person, as I don't want to take up other people's time… I try and do stuff myself''} (P1).}

\revision{Besides toolkit expert, some participants also used A11yMaker AI as a general-knowledge expert, providing suggestions on visual information that they may not have experience with. For example, some participants (P1, P2, P9) weren't familiar with what visual-based thresholds might make sense when setting up their DIY-ATs, such as distance and light intensity, and they found it useful to just ask AI ``what value would make the most sense in this setting?''}

\subsubsection{AI as brainstorming partner and idea starter}
% \ben{I wanted to also present findings on the theme of "Participants found the assistant helpful for narrowing problems and mapping them to the toolkit...", but we don't have any good evidence of participants starting really broad and narrowing down their problem.}
\revision{We also saw evidence of how A11yMaker AI acted as a bridge between participants' layperson needs and technical toolkit setup.  Some participants reflected on how they didn't initially know where to start with the complex toolkit for a certain need they had in mind (P1, P3, P4).
% P1 wanted to build a DIY-AT that could help her "find the c route bus." The 
For example, P1 reported that they sometimes forgot to blow out their candles (due to low light sensitivity) and wanted to build a DIY-AT that could help, but initially weren't sure how to address this using the toolkit. So P1 started by mentioning their challenges to A11yMaker AI and followed the first solution suggested by A11yMaker AI to create a solution: a DIY-AT that generated vibration feedback when a light module still detected light from the candle while the room exiting motion was detected by a motion module, indicating that P1 had left the room but forgot to blow out the candle. P1 later reflected,
\textit{``I kind of have trouble like I don’t know if conceptualizing it is the right [word], but like putting it into practical uses...that’s the nice thing about the [AI], though, is like here’s what I’m having difficulty with, `what do you suggest?' It’s like a team''} (P1). P3 further highlighted the complexity of the toolkit and how the AI assitant abstracting away this complexity was especially useful for a lot of their fellow senior friends who were also not technically savvy.}

\subsection{\revision{Misunderstanding between AI and PVI}}  

\revision{Besides the benefits, we observed multiple cases where AI suggested misleading solutions, highlighting AI barriers when co-making with PVI. We elaborate four major reasons that led to the misunderstanding between AI and PVI.} %Across sessions, the AI assistant often proposed partial or incomplete solutions without clarifying toolkit limitations, which misled participants into believing a task could be fully supported by the toolkit when it could not. In these moments, the AI did not communicate what the toolkit could and could not do. When participants asked it to support tasks beyond what was feasible, the AI often offered vague or misleading guidance that appeared workable but left critical technical gaps. 

\subsubsection{AI overlooking toolkit limitation}  
\revision{%There were times when participants wanted to use the toolkit to create tangible DIY-AT that weren't fully supported by the toolkit. The long tail of DIY-AT PVI might want to build and personalize is long \cite{herskovitz2023hacking}, and we imagine that there will always be use cases that will fall outside the limitations of any toolkit. 
Participants (\textit{e.g.,} P1, P2, P4) sometimes asked for solutions that could not be supported by the tangible toolkit. However, instead of proactively communicating the toolkit capability, we observed that A11yMaker AI often presented problematic solutions that cannot be fulfilled, leaving participants to discover shortcomings later by themselves.}
% A recurring issue was the AI failing to proactively surface the limits of its own suggestions. Instead of flagging gaps, it often presented partial solutions as if they were sufficient, leaving participants to discover shortcomings only after pressing further or drawing on prior experience.  
For instance, P1 asked for help locating a bus stop. While the toolkit supported recognition of humans and certain objects, it could not recognize a bus stop or a bus stop sign. However, the AI assitant proposed ``setting up the camera module to detect any kind of sign, which might help in identifying the bus stop sign if it has recognizable features.''
% \yuhang{better distinguish this from hallucination, or change to another example where AI did not communicate sufficient information instead of hallucinating a function.} 
%accepted this suggestion, but only realized after trying it that it had obvious holes: it could mistake any sign for a bus stop, and it provided no way to guide them once the sign was detected. P1 might have recognized these issues earlier if the AI had proactively raised them, instead of leaving the burden on the participant to discover the gaps.

While some participants accepted the problematic solutions, others challenged the AI assistant based on their prior experience. For example, P2 asked for locating the toilet, but the AI assistant only vaguely suggested using the camera to detect the toilet. P2 further pressed \textit{``How would you determine if I’m aiming correctly or not?''} and the AI assistant then made it clear that the suggestion was not workable. This indicated that the current AI assistant only reacted to limitations passively instead of proactively communicating them with the user. 
% \yuhang{revist... some issue with this example.}

\subsubsection{\revision{AI oversimplifying PVI's needs.}} \revision{The assistant sometimes oversimplified a problem without considering PVI's unique needs.} For example, P3, P6, and P9 all needed help in finding specific items in their environments. In response, the AI assitant suggested using the camera module to detect that object, for example, ``detect things like outlets using visual recognition'' (suggested by A11yMaker AI to P9). However, \revision{it overlooked that PVI may need additional instructions on where the object is and how to reach it beyond the existence of the object in the camera frame.} % clarified that the toolkit could only detect the existence of an object but not where it was or how to reach it. %In these cases, participants moved forward assuming the solution was adequate, when they might have recognized its limits earlier had the AI surfaced them.  

Interestingly, despite AI proposing imperfect solutions, we observed that some participants proactively question AI's suggestion to refine the solution. For example, P4 wanted to identify the ``Badger Bus,'' a specific greyhound bus that is labeled ``Badger Bus'' in big letters on the side of the bus. In response to the AI suggestion that simply alerting P4 when the camera module reads ``Badger Bus,'' P4 brought up an important edge case---what if they stand too close to the bus or at the wrong angle: \textit{``This is a very big bus, big words. I would have to stand pretty far away for you to be able to scan it. It’s not like scanning a sheet of paper''} (P4). Only after P4 surfaced this issue did the AI acknowledge that text detection may struggle with very large moving text and suggested alternatives, such as scanning text near the bus door.

This exchange highlights how the AI assistant failed to anticipate obvious limitations from a non-visual perspective \revision{and how PVI's prior experience and feedback can help further improve the AI suggestions, emphasizing the importance of the collaborative efforts from AI and PVI}. 

% it was only because P4 drew on their prior misfitting experiences with smartphone-based VAT that the issue came to light and was addressed.

\subsubsection{AI hallucinating toolkit capabilities.}  
The AI assistant sometimes hallucinated toolkit capabilities---presenting features that did not exist as if they were real---which left participants confused about why their solutions failed (P1, P7, P9).  
For example, while brainstorming with P1 about detecting bus stop signs, the AI assistant mistakenly suggested that P1 could train the camera module to recognize the signs, even though this was not part of the toolkit’s pretrained classes. In another case, the AI assistant told P9 that it would \textit{``set up the toolkit to receive periodic updates from the temperature module through the sound module,''} even though the sound module could not verbally report temperature readings. % and the system did not support recurring updates. 
%P9 received no confirmation that the setup had worked, and was left unsure why nothing happened.  

Hallucinations also appeared during use. P7 co-made a DIY-AT to detect a dropped water bottle, which triggered a sound alert when the bottle was in the camera module's view. After detection, P7 asked, \textit{``I think we are seeing the water bottle. Where is it exactly?''} and the AI assistant responded with \textit{``To help pinpoint the exact location, try moving the camera module slowly in the direction where the sound is strongest,''} hallucinating that the sound module was attached to the bottle. %a fabricated capability that misled P7 about what the system could do. 
%\yuhang{did AI think that the sound component was attached to the bottle?} 

These cases illustrate that abstraction can backfire when errors are hidden. While lowering technical barriers is important, abstraction should not come at the cost of transparency. PVI need ways to judge for themselves when the AI assistant is hallucinating or when the toolkit is misconfigured, rather than being left unaware of why their solutions fail.

\subsubsection{AI lacking proactive actions to resolve ambiguity}  
Another recurring issue was that the AI assistant failed to clarify ambiguous input from participants. When participants used vague or unconventional wording, the AI assistant often interpreted it literally rather than probing for clarification, which led to irrelevant or confusing responses (P1, P2, P5, P6). For example, P5 repeatedly asked about ``detaching'' the LED module to substitute it with different feedback. The AI assistant treated this as physically removing the module and replied with instructions to unplugging it, rather than recognizing that as swapping feedback modalities. Due to the lack of questions for clarity by the AI assistant, the conversation stalled, leaving P5 without a usable solution.  
This pattern shows how easily misunderstandings could have been avoided if the AI assistant had proactively checked for user intent---asking simple clarifying questions when input was unclear. %, or flagging when a request might not match toolkit functionality. 
By failing to do so, the AI assistant left participants responsible for repairing breakdowns in mutual understanding.  

% %%%%%%%%%%%%%%%%%%%%%%%%%%%%%%%%%%%%%%%%%%%%%%%%%%%%%%%%%%%%%%%
% %%%%%%%%%%%%%%%%%%%%% SUB SUB SECTION #2 %%%%%%%%%%%%%%%%%%%%%
% %%%%%%%%%%%%%%%%%%%%%%%%%%%%%%%%%%%%%%%%%%%%%%%%%%%%%%%%%%%%%%%
% \subsubsection{\textbf{Need for more spatial/visual support for PVI in the tangible, DIY-AT co-making process}}
\subsection{\revision{Spatial and Visual Scaffolding in Tangible Co-making}}
In the physical, DIY-AT co-making process, we observed instances where PVI struggled to design and set up a solution that fully addressed their needs due to a lack of spatial/visual guidance from the LLM-based conversational assistant. Participants sometimes had difficulty orienting the modules correctly, verifying whether or not the components were active, or installing them without asking the researcher present for sighted assistance.

% Should we mention that even when the model was prompted to "help Blind and Low Vision people build custom assistive technology," there were still...

\subsubsection{AI not supporting properly placing and orienting the physical modules}  
% P5, P6, P7, P8, but P5 had a less interesting examples (struggled with no support for orienting the camera modules
% \ben{Maybe sneak in the P1 candle example}
% One area that we observed participants lacking support in, and thus struggled with, was properly placing and orienting the modules in the DIY-AT they created (P5, P6, P7, P8). 
The AI assistant did not provide adequate support for placing and orienting modules, leaving participants unsure how to correctly set up components during co-making (P5, P6, P7, P8). For instance, P6 wanted to use the motion module to detect whether their bag had been moved. The AI assistant only suggested \textit{``Let’s position [the motion module] with the bump facing your bag to detect any movement,''} but P6 still needed sighted confirmation from the researcher to ensure that the placement was correct. P8 also experienced orientation difficulties, initially holding the distance module backward while attempting to detect obstacles, and required correction from the researcher. During the same task, P8 also assumed she could place the distance sensor in her pocket, not realizing this would obstruct it, until the researcher explained that the sensor needs a clear line of sight to detect obstacles. Even P7, who did not misplace a module, was initially unsure how to hold the camera without covering the lens. 

As highlighted by P6 and P8, participants wanted the AI assistant to act more like a sighted guide: \textit{``If you were holding it the wrong way..., the AI assistant wouldn't even know if you were holding it the wrong way''} (P8); \textit{``it would be really nice if the AI assistant could see if it's the right way and let you know''} (P6). These orientation challenges highlight the need for AI systems to provide accessible feedback during installation—for example, by verifying that modules are properly oriented, alerting users when sensors are obstructed, and otherwise supporting steps that would typically rely on vision.

\subsubsection{AI lacking proactive, visual awareness of the toolkit states}  
Another challenge was that the AI assistant did not provide enough visibility into the state of the toolkit. When modules malfunctioned, such as running out of battery or disconnecting unexpectedly, participants were left guessing what might have gone wrong. For example, when P5 tried to locate a module by triggering a sound, no feedback was produced. P5 was unsure if the module was battery-dead or simply located outside the room and out of range. The AI assistant did not clarify the situation, leaving P5 to troubleshoot blindly. These cases highlight the need for the AI assistant to surface real-time information about module states. Proactive awareness---for example, notifying participants when a module has lost power \textit{v.s.} gone out of range---would have helped them diagnose failures more quickly, rather than leaving them uncertain about whether the toolkit was functioning.

% which becomes problematic when the PVI user can't verify these solutions, either because they might be new to the toolkit and because (in the version of the toolkit we created) gives no feedback when erros occur due to hallucination
% highlighting the need for a balance between abstracting away the technical knowledge needed to program the toolkit while also allowing the PVI user to understand when errors occur due to AI hallucination.

% \subsubsection{Need for the AI to be a more proactive in the brainstorm}
% \subsubsection{Need for the AI to be more proactive in preventing}

% section X
% \subsection{AI Transparency Needed to Prevent False Overconfidence}
% \subsection{Important for Proactive AI to Avoid Misunderstanding}

% \textbf{These should be in Discussion?}
% \subsection{Tangible, DIY-AT Creation is Inherently Spatial and Visual: Important Mechanisms an AI Conversational Agent Should Have to Properly Support PVI}
% - It is not enough to just have a simple, LLM-based conversational agent
% \subsection{Need for more explainable AI to contest errors}

\section{Discussion}
% Having an AI assistant involved in physical DIY-AT creation can significantly enhance the usability and accessibility of physical toolkits for non-technical users, as well as raise the creative potential of physical DIY solutions. However, through observing PVI co-design custom DIY-AT solutions end-to-end (from initial ideation to fully implemented systems), we identified important design considerations necessary for ensuring effective, proactive, and context-aware collaboration between users who are PVI and LLM-based AI.

%Our study asked: (1) How do PVI leverage an LLM-powered conversational agent to co-make physical DIY-AT using a distributed toolkit? and (2) What design considerations emerge for conversational AI in this context? 
We found that AI assistance made physical DIY-AT more approachable: participants successfully brainstormed and implemented 14 custom solutions spanning navigation, object search, and environmental awareness. The AI assistant lowered barriers to entry by abstracting away programming details and serving as an ``always-available toolkit expert.'' 
However, our findings also showed that the AI assistant played too much of a role as a tool or natural language interface, and not enough as a collaborative, creative co-design partner. Participants often accepted incomplete or hallucinated solutions because the AI assistant did not engage them in questioning uncertainty, refining partial ideas, or considering toolkit limitations and spatial/visual constraints \revision{particular to PVI}.

These observations extend prior work in two directions. 
First, they echo findings on misfitting with AI in VAT, where PVI develop strategies to verify and contest AI errors \revision{\cite{alharbi2024misfitting}}. In the context of creating DIY-ATs with physical toolkits, however, misfits carry different stakes: hallucinations and underdeveloped solutions do not just misinform, but derail the process of building physical systems, leaving participants unable to realize the ATs they intend to build. 
Second, \revision{our observations} extend prior systems like ProgramAlly \cite{herskovitz2024programAlly}, which demonstrates how multimodal representations can support PVI in creating visual access programs. Our findings show that, comparing to co-creating DIY-ATs through smartphone-based visual filtering programs, physical DIY-AT creation introduces additional challenges in spatial orientation, module placement, and hardware constraints, which \revision{LLM-based AI assistants} must address to function as effective co-design partners.

% \subsection{Implications for Design}
% \subsubsection{Designing for Imperfect AI}
\subsection{Designing for Imperfect AI}
\revision{Across the examples of miscommunication between AI and PVI in our findings, the current tangible DIY-AT toolkit did not always support the tangible DIY-ATs participants aimed to build (\textit{e.g.,} receiving guidance to locate and reach an object). However, the LLM-based AI assistant did not consistently surface these limitations or at times implied capabilities the toolkit did not have, leading PVI to rely on DIY-AT solutions that ultimately did not function as intended. Addressing this issue is essential for aligning LLM-based AI assistants to support PVI in co-creating tangible DIY-ATs, since we believe that no one single toolkit can realistically cover the full range of capabilities that PVI may wish to build \cite{herskovitz2023hacking}, and AI hallucination are unlikely be fully resolved in the near future \cite{cossio2025comprehensive}.}

Taken together, our findings indicate that the key challenge in designing AI for physical DIY-AT lies not in fully preventing hallucinations, but in supporting PVI in identifying, interrogating, and iterating through them.
Participants often treated the AI assistant’s first suggestion as sufficient, even when it only partially addressed their need (\textit{e.g.,} detecting ``any sign'' as a proxy for locating a bus stop). This aligns with Zhou \textit{et al.}’s critique of linear ``user specifies, AI executes'' workflows, which conflict with the inherently nonlinear nature of the design process \cite{zhou2024codesignAI}. Instead, conversational agents must support iterative clarifications, remixing, and alternative exploration. Mapping the framework from Zhou \textit{et al.} \cite{zhou2024codesignAI} onto our setting highlights three priorities: (1) guiding PVI in refining ambiguous needs, (2) reviewing partial solutions to identify gaps, and (3) exposing misunderstandings through back-and-forth question-and-answer mechanisms.

Designing for imperfection also requires mechanisms to mitigate hallucinations. Participants lacked visibility into whether the AI assistant’s programming suggestions were actually feasible within the toolkit. This suggests opportunities to ground outputs \revision{with program verification}---using verified capability schemas, prompting agents to cross-reference suggestions against documentation, or even leveraging multi-agent cross-checking approaches. Rather than eliminating all errors, which is likely impossible, the design goal aims to make errors detectable, explainable, and recoverable.

\revision{While A11yMaker AI is equipped with knowledge about system capabilities and limitations through prompt engineering, we acknowledge that further prompt engineering will also help in preventing the shortcomings observed in 5.1. However, %as evidenced by prior work, 
it is hard to guarantee that there will be no LLM errors---as evidenced by other explorations of LLM-based AI interfaces for programming IoT devices \cite{al-safi2025vega}}.

% \subsubsection{Balancing Power and Abstraction}
\subsection{Balancing Power and Abstraction}
As Li \textit{et al.} argued, abstractions in design tools both empower and constrain: they give users ``power-to'' act but also exert ``power-over'' by shaping how users can act \cite{li2023beyondArtifact}. Our \revision{LLM-based AI assistant} abstracted low-level programming to enable rapid prototyping, but it also hided details that would have helped PVI notice when the AI hallucinated \revision{or when the tangible toolkit could not fully support the DIY-AT solutions they wished to build---ultimately prohibiting the proposed solution from fully aligning with their goals}.

% REVISION: Insert sentence here that both highlights that tangible toolkits don't always cover the long-tail of possible DIY-AT that PVI might want to build or personalize (and possibly that there is always a possibility for errors?).
\revision{The long tail of DIY-ATs that PVI may want to build and personalize is extensive \cite{herskovitz2023hacking}, and we expect additional use cases beyond the scope of such a tangible DIY-AT toolkit.}
\revision{With the current state of LLMs, logic/programming errors caused by LLMs may exist, as evidenced in our study and other explorations of LLM-based AI interfaces for programming IoT devices \cite{al-safi2025vega}}.
For an AI-assisted physical DIY-AT toolkit, the implication is that \revision{AI assistants} should not only simplify but also give PVI more transparency and control when errors occur or the toolkit has limitations. Future designs should enable users move smoothly cross levels of abstraction—whether by exposing reasoning traces, enabling selective verification of toolkit capabilities, or providing layered representations (as explored in ProgramAlly \cite{herskovitz2024programAlly}), where users can toggle between natural language instructions, block-based logic, and capability schemas.

Designing AI as an equal co-design partner, therefore, requires rebalancing power: not removing complexity altogether, but making it accessible, negotiable, and inspectable, so that PVI can contest, refine, and redirect the AI assistant’s suggestions.
% As Li et al. argue, abstractions in design tools both empower and constrain: they give users “power-to” act but also exert “power-over” by shaping how users can act \cite{li2023beyondArtifact}. Our conversational agent abstracted away low-level programming, enabling rapid prototyping, but simultaneously hid details that could have helped participants realize when the AI hallucinated or the solution being co-designed doesn't fully align with the user's goal or needs.
% For physical DIY-AT, the implication is that conversational agents should not only simplify but also allow movement across levels of abstraction. PVI should be able to drill down into module logic, query why a particular solution was suggested, or switch to alternative representations when conversational interaction fails. Designing AI as an equal co-design partner therefore requires rebalancing power: not removing complexity altogether, but making it accessible, negotiable, and inspectable.

% \subsubsection{The Opportunity for the AI to be a Sighted Guide}
\subsection{The Opportunity for the AI to be a Sighted Guide}
Finally, our study revealed a unique design opportunity for physical co-making: the potential role of the AI assistant as a sighted guide. Participants frequently struggled with module placement and orientation, asking, \textit{``Did I place this right?''}—a question the AI assistant could not answer without visual or spatial awareness. This limitation suggests extending conversational agents with multimodal input (\textit{e.g.,} cameras or AR glasses) so that they can provide iterative, situated guidance during setup and troubleshooting.

Prior work on VAT shows that Blind users already contest and adapt around AI errors in visual tasks. Extending this capacity into physical DIY-AT solutions requires AI that not only provides verbal scaffolding but also situational awareness---for example, noticing when a module is backward, when a module is disconnected or has died, or when a sensor’s placement will not yield reliable results. In short, physical co-making underscores how accessibility needs are not only informational but also embodied, necessitating AI systems that collaborate in both conversation and physical space. \revision{With research moving towards multimodal LLMs with real-time spatial understanding of the environment that the user is situated \cite{Magay2025ALA, mao2025spatiallm}, we envision in the future GAI possibly acting as an always available, sighted toolkit expert and brainstorming partner that can contextualize the tangible DIY-ATs by proactively providing spatial-aware instructions and feedback---for example, \textit{``on the ledge by the door seems like a good place to put the motion sensor while still being close enough to the door.''}}

\subsection{Limitations and Future Work}
Below, we discuss limitations in our study methodology and A11yMaker AI prototype, as well as future directions of work.
\revision{During the Solution-Building phase of our user study, some participants needed help from the researcher brainstorming personal access challenges they wanted to try co-making a tangible DIY-AT for using our toolkit probe. We recognize that conducting the study in a lab may have limited participants’ ability to explore real-world uses of the toolkit, with some participants explicitly remarking that if they could bring the toolkit and AI probe at home, they could better explore using it on personal, every-day challenges.}
% \revision{Additionally, A11yMaker AI supports only a limited set of predefined event types in its current form. Participants frequently envisioned use cases beyond these capabilities, suggesting the need for more flexibility. Future work should explore AI-driven code generation to expand functionality without manually adding new blocks in the algorithm.}
\revision{Another limitation of the current A11yBits system lies in its expressiveness and extensibility. In its current form, A11yMaker AI supports only a limited set of predefined event types, and participants frequently envisioned use cases beyond these capabilities, pointing to the need for greater flexibility. Future work could explore AI-driven code generation to expand system functionality without requiring manual addition of new algorithmic building blocks.}

\revision{Relatedly, although A11yBits is currently designed as a standalone toolkit, participants expressed interest in integrating it with existing smart and IoT devices to further expand creative possibilities. Participants did not view the smartphone as a replacement for existing modules, but rather as a complementary component that preserves the distributed nature of the toolkit. Many PVI already rely on smartphones, smart speakers, and other connected devices in their daily lives, and several participants suggested incorporating the smartphone itself as an additional module due to its rich sensing and feedback capabilities. This is something that future work could explore.}

\section{Conclusion}
%We presented an AI-assisted tangible DIY-AT toolkit probe and reported insights from an exploratory study with nine PVI participants. Our findings highlight both opportunities and challenges in the co-making of tangible DIY-ATs with conversational AI: the need for greater spatial and visual support, strategies for mitigating novel AI errors, and implications for designing more accessible AI-assisted prototyping.
\revision{In this work, we examined how PVI engage with an-LLM-based conversational assistant during the co-making of physical DIY-AT solutions and reported insights from an exploratory study with nine PVI participants. Our findings highlight both opportunities and challenges in the co-making of tangible DIY-ATs with conversational AI: the need for greater spatial and visual support \revision{(\textit{i.e.} multimodal support)}, strategies for mitigating AI errors, and implications for designing more accessible AI-assisted prototyping.}

\begin{acks}
This research, in part, was supported by an NSF Graduate Research Fellowship for Ben Kosa. Thank you to all of our participants and members of the AI-Assisted Vision Mini-Workshop at the University of Wisconsin--Madison \footnote{https://www.cs.wisc.edu/ai-assisted-vision-mini-workshop/} for their valuable insight. Thank you to the members of the MadAbility Lab for their feedback.
\end{acks}

\bibliographystyle{ACM-Reference-Format}
\bibliography{A11yBits_Ref}

@inproceedings{education,
author = {Thomas, Danielle R and Lin, Jionghao and Gatz, Erin and Gurung, Ashish and Gupta, Shivang and Norberg, Kole and Fancsali, Stephen E and Aleven, Vincent and Branstetter, Lee and Brunskill, Emma and Koedinger, Kenneth R},
title = {Improving Student Learning with Hybrid Human-AI Tutoring: A Three-Study Quasi-Experimental Investigation},
year = {2024},
isbn = {9798400716188},
publisher = {Association for Computing Machinery},
address = {New York, NY, USA},
url = {https://doi.org/10.1145/3636555.3636896},
doi = {10.1145/3636555.3636896},
booktitle = {Proceedings of the 14th Learning Analytics and Knowledge Conference},
pages = {404–415},
numpages = {12},
location = {Kyoto, Japan},
series = {LAK '24}
}

@inproceedings{misfittingwithAI,
author = {Alharbi, Rahaf and Lor, Pa and Herskovitz, Jaylin and Schoenebeck, Sarita and Brewer, Robin N.},
title = {Misfitting With AI: How Blind People Verify and Contest AI Errors},
year = {2024},
isbn = {9798400706776},
publisher = {Association for Computing Machinery},
address = {New York, NY, USA},
url = {https://doi.org/10.1145/3663548.3675659},
doi = {10.1145/3663548.3675659},
booktitle = {Proceedings of the 26th International ACM SIGACCESS Conference on Computers and Accessibility},
articleno = {61},
numpages = {17},
location = {St. John's, NL, Canada},
series = {ASSETS '24}
}

@inproceedings{Sloan2010ThePO,
  title={The potential of adaptive interfaces as an accessibility aid for older web users},
  author={David Sloan and Matthew T. Atkinson and Colin H. C. Machin and Yunqiu Li},
  booktitle={International Cross-Disciplinary Conference on Web Accessibility},
  year={2010},
  url={https://api.semanticscholar.org/CorpusID:15617896}
}

@article{garrido2012personalized,
  title={Personalized web accessibility using client-side refactoring},
  author={Garrido, Alejandra and Firmenich, Sergio and Rossi, Gustavo and Grigera, Julian and Medina-Medina, Nuria and Harari, Ivana},
  journal={IEEE Internet Computing},
  volume={17},
  number={4},
  pages={58--66},
  year={2012},
  publisher={IEEE}
}

@inproceedings{martin2018adaptiveWebNavigation,
author = {Martin-Hammond, Aqueasha and Hamidi, Foad and Bhalerao, Tejas and Ortega, Christian and Ali, Abdullah and Hornback, Catherine and Means, Casey and Hurst, Amy},
title = {Designing an Adaptive Web Navigation Interface for Users with Variable Pointing Performance},
year = {2018},
isbn = {9781450356510},
publisher = {Association for Computing Machinery},
address = {New York, NY, USA},
url = {https://doi.org/10.1145/3192714.3192818},
doi = {10.1145/3192714.3192818},
booktitle = {Proceedings of the 15th International Web for All Conference},
articleno = {31},
numpages = {10},
location = {Lyon, France},
series = {W4A '18}
}

@inproceedings{ciolfi2016articulating,
  title={Articulating co-design in museums: Reflections on two participatory processes},
  author={Ciolfi, Luigina and Avram, Gabriela and Maye, Laura and Dulake, Nick and Marshall, Mark T and van Dijk, Dick and McDermott, Fiona},
  booktitle={Proceedings of the 19th ACM conference on computer-supported cooperative work \& social computing},
  pages={13--25},
  year={2016}
}

@article{liu2003towards,
  title={Towards an ‘ideal’approach for concept generation},
  author={Liu, Y-C and Chakrabarti, Amaresh and Bligh, Thomas},
  journal={Design studies},
  volume={24},
  number={4},
  pages={341--355},
  year={2003},
  publisher={Elsevier}
}

@article{Gamage2023WhatDBA,
  title={What do Blind and Low-Vision People Really Want from Assistive Smart Devices? Comparison of the Literature with a Focus Study},
  author={Bhanuka Gamage and Thanh-Toan Do and N. Price and A. Lowery and Kim Marriott},
  journal={Proceedings of the 25th International ACM SIGACCESS Conference on Computers and Accessibility},
  year={2023},
  url={https://api.semanticscholar.org/CorpusId:264307048}
}

@article{HamiltonFletcher2019SoundSightAMA,
  title={SoundSight: a mobile sensory substitution device that sonifies colour, distance, and temperature},
  author={Giles Hamilton-Fletcher and James Alvarez and Marianna Obrist and J. Ward},
  journal={Journal on Multimodal User Interfaces},
  year={2019},
  volume={16},
  pages={107 - 123},
  url={https://api.semanticscholar.org/CorpusId:240594991}
}

@article{wen2024find,
author = {Wen, Linda Yilin and Morrison, Cecily and Grayson, Martin and Faia Marques, Rita and Massiceti, Daniela and Longden, Camilla and Cutrell, Ed},
title = {Find My Things: Personalized Accessibility through Teachable AI for People who are Blind or Low Vision},
year = {2024},
month = {May},
url = {https://www.microsoft.com/en-us/research/publication/find-my-things-personalized-accessibility-through-teachable-ai-for-people-who-are-blind-or-low-vision/},
pages = {403:1-403:6},
journal = {Extended Abstracts of the CHI Conference on Human Factors in Computing Systems},
}

@inproceedings{kacorri2017people,
  title={People with visual impairment training personal object recognizers: Feasibility and challenges},
  author={Kacorri, Hernisa and Kitani, Kris M and Bigham, Jeffrey P and Asakawa, Chieko},
  booktitle={Proceedings of the 2017 CHI Conference on Human Factors in Computing Systems},
  pages={5839--5849},
  year={2017}
}

@inproceedings{stangl2021going,
  title={Going beyond one-size-fits-all image descriptions to satisfy the information wants of people who are blind or have low vision},
  author={Stangl, Abigale and Verma, Nitin and Fleischmann, Kenneth R and Morris, Meredith Ringel and Gurari, Danna},
  booktitle={Proceedings of the 23rd international ACM SIGACCESS conference on computers and accessibility},
  pages={1--15},
  year={2021}
}

@inproceedings{wang2024gazePrompt,
author = {Wang, Ru and Potter, Zach and Ho, Yun and Killough, Daniel and Zeng, Linxiu and Mondal, Sanbrita and Zhao, Yuhang},
title = {GazePrompt: Enhancing Low Vision People's Reading Experience with Gaze-Aware Augmentations},
year = {2024},
isbn = {9798400703300},
publisher = {Association for Computing Machinery},
address = {New York, NY, USA},
url = {https://doi.org/10.1145/3613904.3642878},
doi = {10.1145/3613904.3642878},
booktitle = {Proceedings of the 2024 CHI Conference on Human Factors in Computing Systems},
articleno = {894},
numpages = {17},
location = {Honolulu, HI, USA},
series = {CHI '24}
}

@inproceedings{Magay2025ALA,
  title={A Light and Smart Wearable Platform with Multimodal Foundation Model for Enhanced Spatial Reasoning in People with Blindness and Low Vision},
  author={Alexey Magay and Dhurba Tripathi and Yu Hao and Yi Fang},
  booktitle={ECCV Workshops},
  year={2025},
  url={https://api.semanticscholar.org/CorpusID:278714599}
}

@article{mao2025spatiallm,
  title={SpatialLM: Training Large Language Models for Structured Indoor Modeling},
  author={Mao, Yongsen and Zhong, Junhao and Fang, Chuan and Zheng, Jia and Tang, Rui and Zhu, Hao and Tan, Ping and Zhou, Zihan},
  journal={arXiv preprint arXiv:2506.07491},
  year={2025}
}

@inproceedings{Amershi2019Guidelines,
  author       = {Amershi, Saleema and Weld, Dan and Vorvoreanu, Mihaela and Fourney, Adam and Nushi, Besmira and Collisson, Penny and Suh, Jina and Iqbal, Shamsi and Bennett, Paul N. and Inkpen, Kori and Teevan, Jaime and Kikin-Gil, Ruth and Horvitz, Eric},
  title        = {Guidelines for Human-AI Interaction},
  booktitle    = {Proceedings of the 2019 CHI Conference on Human Factors in Computing Systems},
  year         = {2019},
  pages        = {1–13},
  address      = {Glasgow, Scotland Uk},
  publisher    = {ACM},
  doi          = {10.1145/3290605.3300233}
}

@Article{al-safi2025vega,
AUTHOR = {Al-Safi, Harith and Ibrahim, Harith and Steenson, Paul},
TITLE = {Vega: LLM-Driven Intelligent Chatbot Platform for Internet of Things Control and Development},
JOURNAL = {Sensors},
VOLUME = {25},
YEAR = {2025},
NUMBER = {12},
ARTICLE-NUMBER = {3809},
URL = {https://www.mdpi.com/1424-8220/25/12/3809},
PubMedID = {40573696},
ISSN = {1424-8220},
DOI = {10.3390/s25123809}
}

@inproceedings{glazko2024disabilityBias,
author = {Glazko, Kate and Mohammed, Yusuf and Kosa, Ben and Potluri, Venkatesh and Mankoff, Jennifer},
title = {Identifying and Improving Disability Bias in GPT-Based Resume Screening},
year = {2024},
isbn = {9798400704505},
publisher = {Association for Computing Machinery},
address = {New York, NY, USA},
url = {https://doi.org/10.1145/3630106.3658933},
doi = {10.1145/3630106.3658933},
booktitle = {Proceedings of the 2024 ACM Conference on Fairness, Accountability, and Transparency},
pages = {687–700},
numpages = {14},
location = {Rio de Janeiro, Brazil},
series = {FAccT '24}
}

@inproceedings{gadiraju2023offensive,
author = {Gadiraju, Vinitha and Kane, Shaun and Dev, Sunipa and Taylor, Alex and Wang, Ding and Denton, Remi and Brewer, Robin},
title = {"I wouldn't say offensive but...": Disability-Centered Perspectives on Large Language Models},
year = {2023},
isbn = {9798400701924},
publisher = {Association for Computing Machinery},
address = {New York, NY, USA},
url = {https://doi.org/10.1145/3593013.3593989},
doi = {10.1145/3593013.3593989},
booktitle = {Proceedings of the 2023 ACM Conference on Fairness, Accountability, and Transparency},
pages = {205–216},
numpages = {12},
location = {Chicago, IL, USA},
series = {FAccT '23}
}

@article{cossio2025comprehensive,
  title={A comprehensive taxonomy of hallucinations in large language models},
  author={Cossio, Manuel},
  journal={arXiv preprint arXiv:2508.01781},
  year={2025}
}

@inproceedings{stefanie2023style2Fab,
author = {Faruqi, Faraz and Katary, Ahmed and Hasic, Tarik and Abdel-Rahman, Amira and Rahman, Nayeemur and Tejedor, Leandra and Leake, Mackenzie and Hofmann, Megan and Mueller, Stefanie},
title = {Style2Fab: Functionality-Aware Segmentation for Fabricating Personalized 3D Models with Generative AI},
year={2023},
isbn = {9798400701320},
publisher = {Association for Computing Machinery},
address = {New York, NY, USA},
url = {https://doi-org.ezproxy.library.wisc.edu/10.1145/3586183.3606723},
doi = {10.1145/3586183.3606723},
booktitle = {Proceedings of the 36th Annual ACM Symposium on User Interface Software and Technology},
articleno = {22},
numpages = {13},
location = {San Francisco, CA, USA},
series = {UIST '23}
}

@inproceedings{aflatoony2025aifabrication,
author = {Aflatoony, Leila and Li, Mixuan and Zhang, Yiyun and Jacob, Irene and Xu, Shujian and Tang, Ziqi and Grossberg, Andre},
title = {Exploring AI-Fabrication in Shaping the Future of DIY-AT Design: Insights from Makers},
year = {2025},
isbn = {9798400706769},
publisher = {Association for Computing Machinery},
address = {New York, NY, USA},
url = {https://doi.org/10.1145/3663547.3746343},
doi = {10.1145/3663547.3746343},
booktitle = {Proceedings of the 27th International ACM SIGACCESS Conference on Computers and Accessibility},
articleno = {6},
numpages = {15},
location = {
},
series = {ASSETS '25}
}

@inproceedings{mankoff2016clinicalMakerPerspectives,
author = {Hofmann, Megan and Burke, Julie and Pearlman, Jon and Fiedler, Goeran and Hess, Andrea and Schull, Jon and Hudson, Scott E. and Mankoff, Jennifer},
title = {Clinical and Maker Perspectives on the Design of Assistive Technology with Rapid Prototyping Technologies},
year = {2016},
isbn = {9781450341240},
publisher = {Association for Computing Machinery},
address = {New York, NY, USA},
url = {https://doi.org/10.1145/2982142.2982181},
doi = {10.1145/2982142.2982181},
booktitle = {Proceedings of the 18th International ACM SIGACCESS Conference on Computers and Accessibility},
pages = {251–256},
numpages = {6},
location = {Reno, Nevada, USA},
series = {ASSETS '16}
}

@inproceedings{kane2015thingverse,
author = {Buehler, Erin and Branham, Stacy and Ali, Abdullah and Chang, Jeremy J. and Hofmann, Megan Kelly and Hurst, Amy and Kane, Shaun K.},
title = {Sharing is Caring: Assistive Technology Designs on Thingiverse},
year = {2015},
isbn = {9781450331456},
publisher = {Association for Computing Machinery},
address = {New York, NY, USA},
url = {https://doi.org/10.1145/2702123.2702525},
doi = {10.1145/2702123.2702525},
booktitle = {Proceedings of the 33rd Annual ACM Conference on Human Factors in Computing Systems},
pages = {525–534},
numpages = {10},
location = {Seoul, Republic of Korea},
series = {CHI '15}
}

@inproceedings{kelly20153dPrintingProsthetic,
author = {Hofmann, Megan Kelly},
title = {Making Connections: Modular 3D Printing for Designing Assistive Attachments to Prosthetic Devices},
year = {2015},
isbn = {9781450334006},
publisher = {Association for Computing Machinery},
address = {New York, NY, USA},
url = {https://doi.org/10.1145/2700648.2811323},
doi = {10.1145/2700648.2811323},
booktitle = {Proceedings of the 17th International ACM SIGACCESS Conference on Computers \& Accessibility},
pages = {353–354},
numpages = {2},
location = {Lisbon, Portugal},
series = {ASSETS '15}
}

@inproceedings{hofmann2014GripFab,
author = {Buehler, Erin and Hurst, Amy and Hofmann, Megan},
title = {Coming to grips: 3D printing for accessibility},
year = {2014},
isbn = {9781450327206},
publisher = {Association for Computing Machinery},
address = {New York, NY, USA},
url = {https://doi-org.ezproxy.library.wisc.edu/10.1145/2661334.2661345},
doi = {10.1145/2661334.2661345},
booktitle = {Proceedings of the 16th International ACM SIGACCESS Conference on Computers \& Accessibility},
pages = {291–292},
numpages = {2},
location = {Rochester, New York, USA},
series = {ASSETS '14}
}

@inproceedings{herskovitz2024programAlly,
author = {Herskovitz, Jaylin and Xu, Andi and Alharbi, Rahaf and Guo, Anhong},
title = {ProgramAlly: Creating Custom Visual Access Programs via Multi-Modal End-User Programming},
year = {2024},
isbn = {9798400706288},
publisher = {Association for Computing Machinery},
address = {New York, NY, USA},
url = {https://doi.org/10.1145/3654777.3676391},
doi = {10.1145/3654777.3676391},
booktitle = {Proceedings of the 37th Annual ACM Symposium on User Interface Software and Technology},
articleno = {85},
numpages = {15},
location = {Pittsburgh, PA, USA},
series = {UIST '24}
}

@inproceedings{zhao2023a11yBits,
author = {He, Liwen and Li, Yifan and Fan, Mingming and He, Liang and Zhao, Yuhang},
title = {A Multi-modal Toolkit to Support DIY Assistive Technology Creation for Blind and Low Vision People},
year = {2023},
isbn = {9798400700965},
publisher = {Association for Computing Machinery},
address = {New York, NY, USA},
url = {https://doi.org/10.1145/3586182.3616646},
doi = {10.1145/3586182.3616646},
booktitle = {Adjunct Proceedings of the 36th Annual ACM Symposium on User Interface Software and Technology},
articleno = {3},
numpages = {3},
location = {San Francisco, CA, USA},
series = {UIST '23 Adjunct}
}

@inproceedings{zhou2024codesignAI,
author = {Zhou, Jiayi and Li, Renzhong and Tang, Junxiu and Tang, Tan and Li, Haotian and Cui, Weiwei and Wu, Yingcai},
title = {Understanding Nonlinear Collaboration between Human and AI Agents: A Co-design Framework for Creative Design},
year = {2024},
isbn = {9798400703300},
publisher = {Association for Computing Machinery},
address = {New York, NY, USA},
url = {https://doi.org/10.1145/3613904.3642812},
doi = {10.1145/3613904.3642812},
booktitle = {Proceedings of the 2024 CHI Conference on Human Factors in Computing Systems},
articleno = {170},
numpages = {16},
location = {Honolulu, HI, USA},
series = {CHI '24}
}

@inproceedings{li2023beyondArtifact,
author = {Li, Jingyi and Rawn, Eric and Ritchie, Jacob and Tran O'Leary, Jasper and Follmer, Sean},
title = {Beyond the Artifact: Power as a Lens for Creativity Support Tools},
year = {2023},
isbn = {9798400701320},
publisher = {Association for Computing Machinery},
address = {New York, NY, USA},
url = {https://doi.org/10.1145/3586183.3606831},
doi = {10.1145/3586183.3606831},
booktitle = {Proceedings of the 36th Annual ACM Symposium on User Interface Software and Technology},
articleno = {47},
numpages = {15},
location = {San Francisco, CA, USA},
series = {UIST '23}
}

@inproceedings{lee2024altcanvas,
  title={AltCanvas: A Tile-Based Editor for Visual Content Creation with Generative AI for Blind or Visually Impaired People},
  author={Lee, Seonghee and Kohga, Maho and Landau, Steve and O'Modhrain, Sile and Subramonyam, Hari},
  booktitle={Proceedings of the 26th International ACM SIGACCESS Conference on Computers and Accessibility},
  pages={1--22},
  year={2024}
}

@inproceedings{chheda2025artinsight,
  title={ArtInsight: Enabling AI-Powered Artwork Engagement for Mixed Visual-Ability Families},
  author={Chheda-Kothary, Arnavi and Kanchi, Ritesh and Sanders, Chris and Xiao, Kevin and Sengupta, Aditya and Kneitmix, Melanie and Wobbrock, Jacob O and Froehlich, Jon E},
  booktitle={Proceedings of the 30th International Conference on Intelligent User Interfaces},
  pages={190--210},
  year={2025}
}

@article{behnke2019alice,
  title={Alice in DIY wonderland or: Instructing novice users on how to use tools in DIY projects},
  author={Behnke, Gregor and Schiller, Marvin and Kraus, Matthias and Bercher, Pascal and Schmautz, Mario and Dorna, Michael and Dambier, Michael and Minker, Wolfgang and Glimm, Birte and Biundo, Susanne},
  journal={AI Communications},
  volume={32},
  number={1},
  pages={31--57},
  year={2019},
  publisher={SAGE Publications Sage UK: London, England}
}

@article{bercher2021yourself,
  title={Do it yourself, but not alone: companion-technology for home improvement—bringing a planning-based interactive DIY assistant to life},
  author={Bercher, Pascal and Behnke, Gregor and Kraus, Matthias and Schiller, Marvin and Manstetten, Dietrich and Dambier, Michael and Dorna, Michael and Minker, Wolfgang and Glimm, Birte and Biundo, Susanne},
  journal={KI-K{\"u}nstliche Intelligenz},
  volume={35},
  number={3},
  pages={367--375},
  year={2021},
  publisher={Springer}
}

@article{chandrasekera2024can,
  title={Can artificial intelligence support creativity in early design processes?},
  author={Chandrasekera, Tilanka and Hosseini, Zahrasadat and Perera, Ubhaya},
  journal={International Journal of Architectural Computing},
  pages={14780771241254637},
  year={2024},
  publisher={SAGE Publications Sage UK: London, England}
}

@article{o2024extending,
  title={Extending human creativity with AI},
  author={O'Toole, Katherine and Horv{\'a}t, Em{\H{o}}ke-{\'A}gnes},
  journal={Journal of Creativity},
  volume={34},
  number={2},
  pages={100080},
  year={2024},
  publisher={Elsevier}
}

@inproceedings{jeon2021fashionq,
  title={FashionQ: an ai-driven creativity support tool for facilitating ideation in fashion design},
  author={Jeon, Youngseung and Jin, Seungwan and Shih, Patrick C and Han, Kyungsik},
  booktitle={Proceedings of the 2021 CHI Conference on Human Factors in Computing Systems},
  pages={1--18},
  year={2021}
}

@inproceedings{li2024exploring,
  title={Exploring the Potential of Generative AI in DIY Assistive Technology Design by Occupational Therapists},
  author={Li, Mixuan and Aflatoony, Leila},
  booktitle={Proceedings of the 26th International ACM SIGACCESS Conference on Computers and Accessibility},
  pages={1--6},
  year={2024}
}

@inproceedings{alharbi2024misfitting,
  title={Misfitting With AI: How Blind People Verify and Contest AI Errors},
  author={Alharbi, Rahaf and Lor, Pa and Herskovitz, Jaylin and Schoenebeck, Sarita and Brewer, Robin N},
  booktitle={Proceedings of the 26th International ACM SIGACCESS Conference on Computers and Accessibility},
  pages={1--17},
  year={2024}
}

@article{wan2024felt,
  title={" It Felt Like Having a Second Mind": Investigating Human-AI Co-creativity in Prewriting with Large Language Models},
  author={Wan, Qian and Hu, Siying and Zhang, Yu and Wang, Piaohong and Wen, Bo and Lu, Zhicong},
  journal={Proceedings of the ACM on Human-Computer Interaction},
  volume={8},
  number={CSCW1},
  pages={1--26},
  year={2024},
  publisher={ACM New York, NY, USA}
}

@inproceedings{bennett2024painting,
  title={Painting with Cameras and Drawing with Text: AI Use in Accessible Creativity},
  author={Bennett, Cynthia L and Shelby, Renee and Rostamzadeh, Negar and Kane, Shaun K},
  booktitle={Proceedings of the 26th International ACM SIGACCESS Conference on Computers and Accessibility},
  pages={1--19},
  year={2024}
}

@inproceedings{dang2025authoring,
  title={Authoring LLM-Based Assistance for Real-World Contexts and Tasks},
  author={Dang, Hai and Lafreniere, Ben and Grossman, Tovi and Todi, Kashyap and Li, Michelle},
  booktitle={Proceedings of the 30th International Conference on Intelligent User Interfaces},
  pages={211--230},
  year={2025}
}

@inproceedings{huh2023genassist,
  title={GenAssist: Making image generation accessible},
  author={Huh, Mina and Peng, Yi-Hao and Pavel, Amy},
  booktitle={Proceedings of the 36th Annual ACM Symposium on User Interface Software and Technology},
  pages={1--17},
  year={2023}
}

@article{ivcevic2024artificial,
  title={Artificial intelligence as a tool for creativity},
  author={Ivcevic, Zorana and Grandinetti, Mike},
  journal={Journal of Creativity},
  volume={34},
  number={2},
  pages={100079},
  year={2024},
  publisher={Elsevier}
}

@inproceedings{huh24designchecker,
author = {Huh, Mina and Pavel, Amy},
title = {DesignChecker: Visual Design Support for Blind and Low Vision Web Developers},
year = {2024},
isbn = {9798400706288},
publisher = {Association for Computing Machinery},
address = {New York, NY, USA},
url = {https://doi.org/10.1145/3654777.3676369},
doi = {10.1145/3654777.3676369},
booktitle = {Proceedings of the 37th Annual ACM Symposium on User Interface Software and Technology},
articleno = {142},
numpages = {19},
location = {Pittsburgh, PA, USA},
series = {UIST '24}
}

@inproceedings{zhang25a11yshape,
author = {Zhang, Zhuohao (Jerry) and Li, Haichang and Yu, Chun Meng and Faruqi, Faraz and Xie, Junan and Kim, Gene S-H and Fan, Mingming and Forbes, Angus and Wobbrock, Jacob O. and Guo, Anhong and He, Liang},
title = {A11yShape: AI-Assisted 3-D Modeling for Blind and Low-Vision Programmers},
year = {2025},
isbn = {9798400706769},
publisher = {Association for Computing Machinery},
address = {New York, NY, USA},
url = {https://doi.org/10.1145/3663547.3746362},
doi = {10.1145/3663547.3746362},
booktitle = {Proceedings of the 27th International ACM SIGACCESS Conference on Computers and Accessibility},
articleno = {84},
numpages = {20},
location = {
},
series = {ASSETS '25}
}

@inproceedings{minoli2024blvaiproductivity,
author = {Perera, Minoli},
title = {Enhancing Productivity Applications for People who are Blind using AI Assistants},
year = {2024},
isbn = {9798400703317},
publisher = {Association for Computing Machinery},
address = {New York, NY, USA},
url = {https://doi.org/10.1145/3613905.3638180},
doi = {10.1145/3613905.3638180},
booktitle = {Extended Abstracts of the CHI Conference on Human Factors in Computing Systems},
articleno = {432},
numpages = {6},
location = {Honolulu, HI, USA},
series = {CHI EA '24}
}

@inproceedings{adnin2024look,
  title={"I look at it as the king of knowledge": How Blind People Use and Understand Generative AI Tools},
  author={Adnin, Rudaiba and Das, Maitraye},
  booktitle={Proceedings of the 26th International ACM SIGACCESS Conference on Computers and Accessibility},
  pages={1--14},
  year={2024}
}

@misc{BeMyAI,
  author       = {Be My Eyes},
  title        = {Introducing: Be My AI},
  howpublished = {\url{https://www.bemyeyes.com/blog/introducing-be-my-ai/}},
  year         = {2023},
  note         = {Accessed: 2025-04-10}
}

@misc{SeeingAI,
  author       = {Microsoft},
  title        = {Seeing AI},
  howpublished = {\url{https://www.microsoft.com/en-us/garage/wall-of-fame/seeing-ai/}},
  year         = {2017},
  note         = {Accessed: 2025-04-10}
}

@inproceedings{glazko2023autoethnographic,
  title={An autoethnographic case study of generative artificial intelligence's utility for accessibility},
  author={Glazko, Kate S and Yamagami, Momona and Desai, Aashaka and Mack, Kelly Avery and Potluri, Venkatesh and Xu, Xuhai and Mankoff, Jennifer},
  booktitle={Proceedings of the 25th International ACM SIGACCESS Conference on Computers and Accessibility},
  pages={1--8},
  year={2023}
}

@article{penuela2025towards,
  title={Towards Understanding the Use of MLLM-Enabled Applications for Visual Interpretation by Blind and Low Vision People},
  author={Penuela, Ricardo E Gonzalez and Hu, Ruiying and Lin, Sharon and Shende, Tanisha and Azenkot, Shiri},
  journal={arXiv preprint arXiv:2503.05899},
  year={2025}
}

@article{mukhiddinov2022automatic,
  title={Automatic fire detection and notification system based on improved YOLOv4 for the blind and visually impaired},
  author={Mukhiddinov, Mukhriddin and Abdusalomov, Akmalbek Bobomirzaevich and Cho, Jinsoo},
  journal={Sensors},
  volume={22},
  number={9},
  pages={3307},
  year={2022},
  publisher={MDPI}
}

@inproceedings{morrison2023understanding,
  title={Understanding personalized accessibility through teachable ai: designing and evaluating find my things for people who are blind or low vision},
  author={Morrison, Cecily and Grayson, Martin and Marques, Rita Faia and Massiceti, Daniela and Longden, Camilla and Wen, Linda and Cutrell, Edward},
  booktitle={Proceedings of the 25th International ACM SIGACCESS Conference on Computers and Accessibility},
  pages={1--12},
  year={2023}
}

@inproceedings{hong2022blind,
  title={Blind users accessing their training images in teachable object recognizers},
  author={Hong, Jonggi and Gandhi, Jaina and Mensah, Ernest Essuah and Zeraati, Farnaz Zamiri and Jarjue, Ebrima and Lee, Kyungjun and Kacorri, Hernisa},
  booktitle={Proceedings of the 24th International ACM SIGACCESS Conference on Computers and Accessibility},
  pages={1--18},
  year={2022}
}

@inproceedings{gonzalez2024investigating,
  title={Investigating use cases of ai-powered scene description applications for blind and low vision people},
  author={Gonzalez Penuela, Ricardo E and Collins, Jazmin and Bennett, Cynthia and Azenkot, Shiri},
  booktitle={Proceedings of the 2024 CHI Conference on Human Factors in Computing Systems},
  pages={1--21},
  year={2024}
}

@inproceedings{ahmetovic2020recog,
  title={Recog: Supporting blind people in recognizing personal objects},
  author={Ahmetovic, Dragan and Sato, Daisuke and Oh, Uran and Ishihara, Tatsuya and Kitani, Kris and Asakawa, Chieko},
  booktitle={Proceedings of the 2020 CHI Conference on Human Factors in Computing Systems},
  pages={1--12},
  year={2020}
}

@inproceedings{gao2024assistgpt,
  title={AssistGPT: Towards Multi-modal Agent for Human-Centric AI Assistant},
  author={Gao, Difei and Hu, Siyuan and Lin, Qinghong and Shou, Mike Zheng},
  booktitle={Proceedings of the 5th International Workshop on Human-centric Multimedia Analysis},
  pages={3--5},
  year={2024}
}

@inproceedings{chen2023origami,
  title={Origami Sensei: mixed reality AI-assistant for creative tasks using hands},
  author={Chen, Qiyu and Mishra, Richa and El-Zanfaly, Dina and Kitani, Kris},
  booktitle={Companion Publication of the 2023 ACM Designing Interactive Systems Conference},
  pages={147--151},
  year={2023}
}

@inproceedings{choudhury2023medisage,
  title={MediSage: An AI Assistant for Healthcare via Composition of Neural-Symbolic Reasoning Operators},
  author={Choudhury, Sutanay and Agarwal, Khushbu and Ham, Colby and Tamang, Suzanne},
  booktitle={Companion Proceedings of the ACM Web Conference 2023},
  pages={258--261},
  year={2023}
}

@article{deldari2024auditnet,
  title={AuditNet: A Conversational AI-based Security Assistant},
  author={Deldari, Shohreh and Goudarzi, Mohammad and Joshi, Aditya and Shaghaghi, Arash and Finn, Simon and Salim, Flora D and Jha, Sanjay},
  journal={arXiv preprint arXiv:2407.14116},
  year={2024}
}

@inproceedings{zhang2018stockassistant,
  title={Stockassistant: a stock ai assistant for reliability modeling of stock comments},
  author={Zhang, Chen and Wang, Yijun and Chen, Can and Du, Changying and Yin, Hongzhi and Wang, Hao},
  booktitle={Proceedings of the 24th ACM SIGKDD international conference on knowledge discovery \& data mining},
  pages={2710--2719},
  year={2018}
}

@inproceedings{jetbrains2024ai,
author = {Sokolov, Andrey},
title = {AI Assistant in JetBrains IDE: Insights and Challenges (Invited Talk)},
year = {2024},
isbn = {9798400706851},
publisher = {Association for Computing Machinery},
address = {New York, NY, USA},
url = {https://doi.org/10.1145/3664646.3676273},
doi = {10.1145/3664646.3676273},
booktitle = {Proceedings of the 1st ACM International Conference on AI-Powered Software},
pages = {178},
numpages = {1},
location = {Porto de Galinhas, Brazil},
series = {AIware 2024}
}

@article{herskovitz2024diy,
  title={DIY Assistive Software: End-User Programming for Personalized Assistive Technology},
  author={Herskovitz, Jaylin},
  journal={ACM SIGACCESS Accessibility and Computing},
  number={137},
  pages={1--1},
  year={2024},
  publisher={ACM New York, NY, USA}
}

@misc{blind_arduino_project,
  author       = {Joshua Miele},
  title        = {The Blind Arduino Project},
  year         = {2016},
  url          = {https://www.ski.org/projects/blind-arduino-project},
  note         = {Accessed: 2025-04-09}
}

@article{zhao2024vialm,
  title={Vialm: A survey and benchmark of visually impaired assistance with large models},
  author={Zhao, Yi and Zhang, Yilin and Xiang, Rong and Li, Jing and Li, Hillming},
  journal={arXiv preprint arXiv:2402.01735},
  year={2024}
}

@inproceedings{yang2024viassist,
  title={Viassist: Adapting multi-modal large language models for users with visual impairments},
  author={Yang, Bufang and He, Lixing and Liu, Kaiwei and Yan, Zhenyu},
  booktitle={2024 IEEE International Workshop on Foundation Models for Cyber-Physical Systems \& Internet of Things (FMSys)},
  pages={32--37},
  year={2024},
  organization={IEEE}
}

@article{phillips1993predictors,
  title={Predictors of assistive technology abandonment},
  author={Phillips, Betsy and Zhao, Hongxin},
  journal={Assistive technology},
  volume={5},
  number={1},
  pages={36--45},
  year={1993},
  publisher={Taylor \& Francis}
}

@inproceedings{lefeuvre2016loaded,
  title={Loaded dice: exploring the design space of connected devices with blind and visually impaired people},
  author={Lefeuvre, Kevin and Totzauer, S{\"o}ren and Bischof, Andreas and Kurze, Albrecht and Storz, Michael and Ullmann, Lisa and Berger, Arne},
  booktitle={Proceedings of the 9th Nordic Conference on Human-Computer Interaction},
  pages={1--10},
  year={2016}
}

@inproceedings{ducasse2016tangible,
  title={Tangible reels: construction and exploration of tangible maps by visually impaired users},
  author={Ducasse, Julie and Mac{\'e}, Marc JM and Serrano, Marcos and Jouffrais, Christophe},
  booktitle={Proceedings of the 2016 CHI conference on human factors in computing systems},
  pages={2186--2197},
  year={2016}
}

@article{steinmetz2021causes,
  title={Causes of blindness and vision impairment in 2020 and trends over 30 years, and prevalence of avoidable blindness in relation to VISION 2020: the Right to Sight: an analysis for the Global Burden of Disease Study},
  author={Steinmetz, Jaimie D and Bourne, Rupert RA and Briant, Paul Svitil and Flaxman, Seth R and Taylor, Hugh RB and Jonas, Jost B and Abdoli, Amir Aberhe and Abrha, Woldu Aberhe and Abualhasan, Ahmed and Abu-Gharbieh, Eman Girum and others},
  journal={The Lancet Global Health},
  volume={9},
  number={2},
  pages={e144--e160},
  year={2021},
  publisher={Elsevier}
}

@inproceedings{brady2013visual,
  title={Visual challenges in the everyday lives of blind people},
  author={Brady, Erin and Morris, Meredith Ringel and Zhong, Yu and White, Samuel and Bigham, Jeffrey P},
  booktitle={Proceedings of the SIGCHI conference on human factors in computing systems},
  pages={2117--2126},
  year={2013}
}

@inproceedings{mennicken2014today,
  title={From today's augmented houses to tomorrow's smart homes: new directions for home automation research},
  author={Mennicken, Sarah and Vermeulen, Jo and Huang, Elaine M},
  booktitle={Proceedings of the 2014 ACM international joint conference on pervasive and ubiquitous computing},
  pages={105--115},
  year={2014}
}

@article{pape2002shaping,
  title={The shaping of individual meanings assigned to assistive technology: a review of personal factors},
  author={Pape, T Louise-Bender and Kim, J and Weiner, B},
  journal={Disability and rehabilitation},
  volume={24},
  number={1-3},
  pages={5--20},
  year={2002},
  publisher={Taylor \& Francis}
}

@misc{WHO,
title={Blindness and vision impairment},
author={World Health Organization},
year={2022},
note={last accessed 8 Nov 2022},
howpublished={\url{https://www.who.int/news-room/fact-sheets/detail/blindness-and-visual-impairment}}
}

@inproceedings{williams2013pray,
  title={" Pray before you step out" describing personal and situational blind navigation behaviors},
  author={Williams, Michele A and Hurst, Amy and Kane, Shaun K},
  booktitle={Proceedings of the 15th International ACM SIGACCESS Conference on Computers and Accessibility},
  pages={1--8},
  year={2013}
}

@inproceedings{zhao2018looks,
  title={" It Looks Beautiful but Scary" How Low Vision People Navigate Stairs and Other Surface Level Changes},
  author={Zhao, Yuhang and Kupferstein, Elizabeth and Tal, Doron and Azenkot, Shiri},
  booktitle={Proceedings of the 20th International ACM SIGACCESS Conference on Computers and Accessibility},
  pages={307--320},
  year={2018}
}

@inproceedings{li2021non,
  title={Non-Visual Cooking: Exploring Practices and Challenges of Meal Preparation by People with Visual Impairments},
  author={Li, Franklin Mingzhe and Dorst, Jamie and Cederberg, Peter and Carrington, Patrick},
  booktitle={The 23rd International ACM SIGACCESS Conference on Computers and Accessibility},
  pages={1--11},
  year={2021}
}

@inproceedings{szpiro2016finding,
  title={Finding a store, searching for a product: a study of daily challenges of low vision people},
  author={Szpiro, Sarit and Zhao, Yuhang and Azenkot, Shiri},
  booktitle={Proceedings of the 2016 ACM International Joint Conference on Pervasive and Ubiquitous Computing},
  pages={61--72},
  year={2016}
}

@inproceedings{zhao2018face,
  title={A face recognition application for people with visual impairments: Understanding use beyond the lab},
  author={Zhao, Yuhang and Wu, Shaomei and Reynolds, Lindsay and Azenkot, Shiri},
  booktitle={Proceedings of the 2018 CHI Conference on Human Factors in Computing Systems},
  pages={1--14},
  year={2018}
}

@inproceedings{jayant2011supporting,
  title={Supporting blind photography},
  author={Jayant, Chandrika and Ji, Hanjie and White, Samuel and Bigham, Jeffrey P},
  booktitle={The proceedings of the 13th international ACM SIGACCESS conference on Computers and accessibility},
  pages={203--210},
  year={2011}
}

@inproceedings{zhao2015foresee,
  title={Foresee: A customizable head-mounted vision enhancement system for people with low vision},
  author={Zhao, Yuhang and Szpiro, Sarit and Azenkot, Shiri},
  booktitle={Proceedings of the 17th international ACM SIGACCESS conference on computers \& accessibility},
  pages={239--249},
  year={2015}
}

@inproceedings{ahmetovic2016navcog,
  title={NavCog: a navigational cognitive assistant for the blind},
  author={Ahmetovic, Dragan and Gleason, Cole and Ruan, Chengxiong and Kitani, Kris and Takagi, Hironobu and Asakawa, Chieko},
  booktitle={Proceedings of the 18th International Conference on Human-Computer Interaction with Mobile Devices and Services},
  pages={90--99},
  year={2016}
}

@inproceedings{bigham2010vizwiz,
  title={Vizwiz: nearly real-time answers to visual questions},
  author={Bigham, Jeffrey P and Jayant, Chandrika and Ji, Hanjie and Little, Greg and Miller, Andrew and Miller, Robert C and Miller, Robin and Tatarowicz, Aubrey and White, Brandyn and White, Samual and others},
  booktitle={Proceedings of the 23nd annual ACM symposium on User interface software and technology},
  pages={333--342},
  year={2010}
}

@inproceedings{zhao2020effectiveness,
  title={The effectiveness of visual and audio wayfinding guidance on smartglasses for people with low vision},
  author={Zhao, Yuhang and Kupferstein, Elizabeth and Rojnirun, Hathaitorn and Findlater, Leah and Azenkot, Shiri},
  booktitle={Proceedings of the 2020 CHI conference on human factors in computing systems},
  pages={1--14},
  year={2020}
}

@article{abras2004user,
  title={User-centered design},
  author={Abras, Chadia and Maloney-Krichmar, Diane and Preece, Jenny and others},
  journal={Bainbridge, W. Encyclopedia of Human-Computer Interaction. Thousand Oaks: Sage Publications},
  volume={37},
  number={4},
  pages={445--456},
  year={2004}
}

@incollection{okerlund2019diy,
  title={DIY assistive technology for others: Considering social impacts and opportunities to leverage HCI techniques},
  author={Okerlund, Johanna and Wilson, David},
  booktitle={Proceedings of FabLearn 2019},
  pages={152--155},
  year={2019}
}

@inproceedings{hook2014study,
  title={A study of the challenges related to DIY assistive technology in the context of children with disabilities},
  author={Hook, Jonathan and Verbaan, Sanne and Durrant, Abigail and Olivier, Patrick and Wright, Peter},
  booktitle={Proceedings of the 2014 conference on Designing interactive systems},
  pages={597--606},
  year={2014}
}

@inproceedings{hurst2013making,
  title={Making" making" accessible},
  author={Hurst, Amy and Kane, Shaun},
  booktitle={Proceedings of the 12th international conference on interaction design and children},
  pages={635--638},
  year={2013}
}

@article{scherer2002change,
  title={The change in emphasis from people to person: introduction to the special issue on Assistive Technology},
  author={Scherer, Marcia J},
  journal={Disability and rehabilitation},
  volume={24},
  number={1-3},
  pages={1--4},
  year={2002},
  publisher={Taylor \& Francis}
}

@inproceedings{sato2017navcog3,
  title={Navcog3: An evaluation of a smartphone-based blind indoor navigation assistant with semantic features in a large-scale environment},
  author={Sato, Daisuke and Oh, Uran and Naito, Kakuya and Takagi, Hironobu and Kitani, Kris and Asakawa, Chieko},
  booktitle={Proceedings of the 19th International ACM SIGACCESS Conference on Computers and Accessibility},
  pages={270--279},
  year={2017}
}

@inproceedings{hurst2011empowering,
  title={Empowering individuals with do-it-yourself assistive technology},
  author={Hurst, Amy and Tobias, Jasmine},
  booktitle={The proceedings of the 13th international ACM SIGACCESS conference on Computers and accessibility},
  pages={11--18},
  year={2011}
}

@article{verza2006interdisciplinary,
  title={An interdisciplinary approach to evaluating the need for assistive technology reduces equipment abandonment},
  author={Verza, Riccardo and Carvalho, ML Lopes and Battaglia, Mario A and Uccelli, M Messmer},
  journal={Multiple Sclerosis Journal},
  volume={12},
  number={1},
  pages={88--93},
  year={2006},
  publisher={Sage Publications Sage CA: Thousand Oaks, CA}
}

@article{copley2004barriers,
  title={Barriers to the use of assistive technology for children with multiple disabilities},
  author={Copley, Jodie and Ziviani, Jenny},
  journal={Occupational therapy international},
  volume={11},
  number={4},
  pages={229--243},
  year={2004},
  publisher={Wiley Online Library}
}

@inproceedings{brown2012viztouch,
  title={VizTouch: automatically generated tactile visualizations of coordinate spaces},
  author={Brown, Craig and Hurst, Amy},
  booktitle={Proceedings of the Sixth International Conference on Tangible, Embedded and Embodied Interaction},
  pages={131--138},
  year={2012}
}

@inproceedings{giles2015imagining,
  title={Imagining future technologies: eTextile weaving workshops with blind and visually impaired people},
  author={Giles, Emilie and Van Der Linden, Janet},
  booktitle={Proceedings of the 2015 ACM SIGCHI Conference on Creativity and Cognition},
  pages={3--12},
  year={2015}
}

@inproceedings{taylor2016making,
  title={Making community: the wider role of makerspaces in public life},
  author={Taylor, Nick and Hurley, Ursula and Connolly, Philip},
  booktitle={Proceedings of the 2016 CHI Conference on human factors in Computing systems},
  pages={1415--1425},
  year={2016}
}

@inproceedings{kuznetsov2010rise,
  title={Rise of the expert amateur: DIY projects, communities, and cultures},
  author={Kuznetsov, Stacey and Paulos, Eric},
  booktitle={Proceedings of the 6th Nordic conference on human-computer interaction: extending boundaries},
  pages={295--304},
  year={2010}
}

@article{Chen2022LiSee,
author = {Chen, Kaixin and Huang, Yongzhi and Chen, Yicong and Zhong, Haobin and Lin, Lihua and Wang, Lu and Wu, Kaishun},
title = {LiSee: A Headphone That Provides All-Day Assistance for Blind and Low-Vision Users to Reach Surrounding Objects},
year = {2022},
issue_date = {September 2022},
publisher = {Association for Computing Machinery},
address = {New York, NY, USA},
volume = {6},
number = {3},
url = {https://doi.org/10.1145/3550282},
doi = {10.1145/3550282},
journal = {Proc. ACM Interact. Mob. Wearable Ubiquitous Technol.},
month = {sep},
articleno = {104},
numpages = {30}
}

@article{Boldu2018FingerReader2,
author = {Boldu, Roger and Dancu, Alexandru and Matthies, Denys J.C. and Buddhika, Thisum and Siriwardhana, Shamane and Nanayakkara, Suranga},
title = {FingerReader2.0: Designing and Evaluating a Wearable Finger-Worn Camera to Assist People with Visual Impairments While Shopping},
year = {2018},
issue_date = {September 2018},
publisher = {Association for Computing Machinery},
address = {New York, NY, USA},
volume = {2},
number = {3},
url = {https://doi.org/10.1145/3264904},
doi = {10.1145/3264904},
journal = {Proc. ACM Interact. Mob. Wearable Ubiquitous Technol.},
month = {sep},
articleno = {94},
numpages = {19}
}

@book{saldana2021coding,
  title={The coding manual for qualitative researchers},
  author={Salda{\~n}a, Johnny},
  year={2021},
  publisher={sage}
}

@article{braun2006using,
  title={Using thematic analysis in psychology},
  author={Braun, Virginia and Clarke, Victoria},
  journal={Qualitative research in psychology},
  volume={3},
  number={2},
  pages={77--101},
  year={2006},
  publisher={Taylor \& Francis}
}

@inproceedings{zhao2016cuesee,
  title={CueSee: exploring visual cues for people with low vision to facilitate a visual search task},
  author={Zhao, Yuhang and Szpiro, Sarit and Knighten, Jonathan and Azenkot, Shiri},
  booktitle={Proceedings of the 2016 ACM International Joint Conference on Pervasive and Ubiquitous Computing},
  pages={73--84},
  year={2016}
}

@inproceedings{huppert2021guidecopter,
  title={GuideCopter-A precise drone-based haptic guidance interface for blind or visually impaired people},
  author={Huppert, Felix and Hoelzl, Gerold and Kranz, Matthias},
  booktitle={Proceedings of the 2021 CHI Conference on Human Factors in Computing Systems},
  pages={1--14},
  year={2021}
}

@inproceedings{laux1996designing,
  title={Designing the World Wide Web for people with disabilities: a user centered design approach},
  author={Laux, Lila F and McNally, Peter R and Paciello, Michael G and Vanderheiden, Gregg C},
  booktitle={Proceedings of the second annual ACM conference on Assistive technologies},
  pages={94--101},
  year={1996}
}

@article{sanchez2008user,
  title={User-Centered Technologies for blind children},
  author={S{\'a}nchez, Jaime},
  journal={Human technology: An Interdisciplinary Journal on Humans in ICT Environments},
  year={2008},
  publisher={University of Jyv{\"a}skyl{\"a}, Agora Center}
}

@inproceedings{ezaki2005improved,
  title={Improved text-detection methods for a camera-based text reading system for blind persons},
  author={Ezaki, Nobuo and Kiyota, Kimiyasu and Minh, Bui Truong and Bulacu, Marius and Schomaker, Lambert},
  booktitle={Eighth International Conference on Document Analysis and Recognition (ICDAR'05)},
  pages={257--261},
  year={2005},
  organization={IEEE}
}

@article{real2019navigation,
  title={Navigation systems for the blind and visually impaired: Past work, challenges, and open problems},
  author={Real, Santiago and Araujo, Alvaro},
  journal={Sensors},
  volume={19},
  number={15},
  pages={3404},
  year={2019},
  publisher={MDPI}
}

@inproceedings{bigham2010vizwizlocate,
  title={VizWiz:: LocateIt-enabling blind people to locate objects in their environment},
  author={Bigham, Jeffrey P and Jayant, Chandrika and Miller, Andrew and White, Brandyn and Yeh, Tom},
  booktitle={2010 IEEE Computer Society Conference on Computer Vision and Pattern Recognition-Workshops},
  pages={65--72},
  year={2010},
  organization={IEEE}
}

@inproceedings{guo2016vizlens,
  title={Vizlens: A robust and interactive screen reader for interfaces in the real world},
  author={Guo, Anhong and Chen, Xiang'Anthony' and Qi, Haoran and White, Samuel and Ghosh, Suman and Asakawa, Chieko and Bigham, Jeffrey P},
  booktitle={Proceedings of the 29th annual symposium on user interface software and technology},
  pages={651--664},
  year={2016}
}

@inproceedings{minatani2019smart,
  title={Smart apps vs. renovated low-tech devices with DIY assistive technology: a case of a banknote identifier for visually impaired people},
  author={Minatani, Kazunori},
  booktitle={Proceedings of the 5th EAI International Conference on Smart Objects and Technologies for Social Good},
  pages={96--101},
  year={2019}
}

@article{herskovitz2023hacking,
  title={Hacking, Switching, Combining: Understanding and Supporting DIY Assistive Technology Design by Blind People},
  author={Herskovitz, Jaylin and Xu, Andi and Alharbi, Rahaf and Guo, Anhong},
  year={2023}
}

@article{kintsch2002framework,
  title={A framework for the adoption of assistive technology},
  author={Kintsch, Anja and DePaula, Rogerio},
  journal={SWAAAC 2002: Supporting learning through assistive technology},
  volume={3},
  pages={1--10},
  year={2002},
  publisher={Citeseer}
}

@book{riemer1997factors,
  title={Factors associated with assistive technology discontinuance among individuals with disabilities},
  author={Riemer-Reiss, Marti Lynn},
  year={1997},
  publisher={University of Northern Colorado}
}

@inproceedings{dawe2006desperately,
  title={Desperately seeking simplicity: how young adults with cognitive disabilities and their families adopt assistive technologies},
  author={Dawe, Melissa},
  booktitle={Proceedings of the SIGCHI conference on Human Factors in computing systems},
  pages={1143--1152},
  year={2006}
}

@inproceedings{shinohara2011shadow,
  title={In the shadow of misperception: assistive technology use and social interactions},
  author={Shinohara, Kristen and Wobbrock, Jacob O},
  booktitle={Proceedings of the SIGCHI conference on human factors in computing systems},
  pages={705--714},
  year={2011}
}

@inproceedings{buehler2015sharing,
  title={Sharing is caring: Assistive technology designs on thingiverse},
  author={Buehler, Erin and Branham, Stacy and Ali, Abdullah and Chang, Jeremy J and Hofmann, Megan Kelly and Hurst, Amy and Kane, Shaun K},
  booktitle={Proceedings of the 33rd Annual ACM Conference on Human Factors in Computing Systems},
  pages={525--534},
  year={2015}
}

@inproceedings{hamidi2018participatory,
  title={Participatory design of DIY digital assistive technology in Western Kenya},
  author={Hamidi, Foad and Mbullo, Patrick and Onyango, Deurence and Hynie, Michaela and McGrath, Susan and Baljko, Melanie},
  booktitle={Proceedings of the Second African Conference for Human Computer Interaction: Thriving Communities},
  pages={1--11},
  year={2018}
}

@inproceedings{meissner2017yourself,
  title={Do-it-yourself empowerment as experienced by novice makers with disabilities},
  author={Meissner, Janis Lena and Vines, John and McLaughlin, Janice and Nappey, Thomas and Maksimova, Jekaterina and Wright, Peter},
  booktitle={Proceedings of the 2017 conference on designing interactive systems},
  pages={1053--1065},
  year={2017}
}

% If your work has an appendix, this is the place to put it.
\appendix

\section{Tangible Module Functionality \& Design}
\label{appendix:tangible}
The Sensing modules include: 
%\liang{ All the module description could follow this format: One sentence to describe what the sensor is and what it can do, followed with the rationale for developing this module and a few application examples that can be used in our context} 

\begin{itemize}
    \item \textit{\textbf{Camera:}} The Camera module captures live visual input from the environment and transmits the video stream to the AI-assisted app via WiFi, where all processing is handled. This module provides PVI access to real-time visual information, allowing them to integrate image-based sensing and recognition into their DIY-ATs.
\textit{Examples of applications include object identification, monitoring activity in a room, and performing text recognition.}

            % \textit{OV5640} 5-megapixel CMOS image sensor
    \item \textbf{\textit{Distance:}} Equipped with a time-of-flight sensor, the Distance module detects the presence of nearby objects within two meters. 
It offers users a way to sense proximity and spatial changes in their environment. 
\textit{Example applications include detecting obstacles in the front when distance changes.}

%supporting various tasks like obstacle avoidance and enhancing spatial awareness of PVI. 
            % The module uses a \textit{VL53L0X} time-of-flight sensor, which emits infrared pulses and calculates the distance based on the time it takes for the signal to return.
    \item \textbf{\textit{Motion:}} 
The Motion module embeds a passive infrared (PIR) sensor that detects motion within its surrounding area. 
This module allows users to respond to movement in their environment. 
\textit{Example applications include detecting people entering a room.}

    % \textit{BS312} This module can be used for presence detection or triggering events based on human movement.
    \item \textbf{\textit{Light:}} The Light module employs a digital light sensor to measure the ambient brightness level and detect changes in environmental lighting conditions.
This module enables users to build solutions that respond to variations in light. 
\textit{Example applications include detecting day-to-night transitions, triggering alerts when in dark environments, or providing cues when a light is turned on or off.}

    % \textit{TSL2561} I\textsuperscript{2}C 
    \item  \textbf{\textit{Posture:}} 
The Posture module can detect how it is moved, tilted, or positioned with a low-power, three-axis accelerometer. %that collects the inertia data along the X-, Y-, and Z- axes, enabling various motion-related applications, such as orientation sensing, gesture/posture recognition,  and tilt and fall detection. 
It was designed to help users sense changes in orientation or motion of an objects or surface it is attached to.
\textit{Example applications include recognizing gestures, detecting when something tips over, or monitoring for sudden movements like falls.}

    % \textit{MMA7660FC}
    \item \textbf{\textit{Temperature:}} The Temperature module detects ambient temperature using a built-in thermistor.
It gives users a way to monitor temperature changes in their environment. 
\textit{Example applications can be alerting users when water gets too hot and monitoring room temperature.}

\end{itemize}

The Feedback modules include:
% PVI have different needs due to various visual abilities. 
% For example, blind people may prefer auditory and haptic notifications, while low-vision people may prefer visual cues \cite{zhao2015foresee, szpiro2016finding}. 
% Prior research shows that appropriate modalities of aids can effectively assist PVI with their daily tasks \cite{potluri2019}. 
% There are four \textit{Feedback Modules} to provide users with multi-modal feedback, including:

\begin{itemize}
    \item \textit{\textbf{Sound:}} The Sound module provides auditory feedback.  
It plays pre-defined sound patterns to communicate different levels of importance in everyday situations. We define three sound patterns:
(i) a \textbf{soft notification} with two short beeps at the same tone, useful for gentle reminders in quiet, non-urgent settings;
(ii) a \textbf{strong notification} with an ascending tone, which naturally draws attention and can signal something is ready or complete; and
(iii) a \textbf{warning sound} using a high-low beep pattern, similar to traditional alarms, intended for urgent or emergency situations.

    % \textit{LM386} 
    \item \textit{\textbf{Vibration:}} The Vibration module provides vibrotactile feedback through a coin vibration motor. % is used to deliver expressive vibration patterns by adjusting the vibration frequency, amplitude, and duration. 
We offer three pre-defined vibration patterns to express different levels of urgency and meaning:
(i) a \textbf{soft notification} with two short, gentle vibrations, suitable for subtle alerts that do not interrupt the user or others nearby;
(ii) a \textbf{strong notification} with one long vibration that gradually increases in strength, helping draw attention to something important or newly activated; and
(iii) a \textbf{warning pattern} with two long, strong vibrations, designed to clearly signal urgent or potentially hazardous events.

    % \textit{ELB060416}
    \item \textit{\textbf{Light Display:}} For people with low vision, the Light Display module provides visual cues through a $6\times10$ LED matrix capable of showing different light colors and patterns. 
We pre-define three light patterns, including
(i) a \textbf{weak notification} using a single yellow light at the center, suitable for long-term, non-urgent events while minimizing disruption to others nearby;
(ii) a \textbf{strong notification} where all green lights turn on, designed to quickly capture the attention of users with low vision; and
(iii) a \textbf{warning pattern} with all red lights flashing, intended to alert the user in more urgent or emergency situations.

    % \textit{WS2812} 
    \item \textit{\textbf{Motor:}} The Motor module provides physical movement. It features a small rotating platform that can change position on its own, allowing users to move or reorient attached objects without handling them manually.
Users can attach other modules or everyday items to the platform to extend functionality.
\textit{For example, attaching the Camera module to scan a wider area via automatically rotating, or attaching it to a light switch to automatically turn it on and off.}

\end{itemize}

\begin{figure}[h]
  \includegraphics[width=1\linewidth]{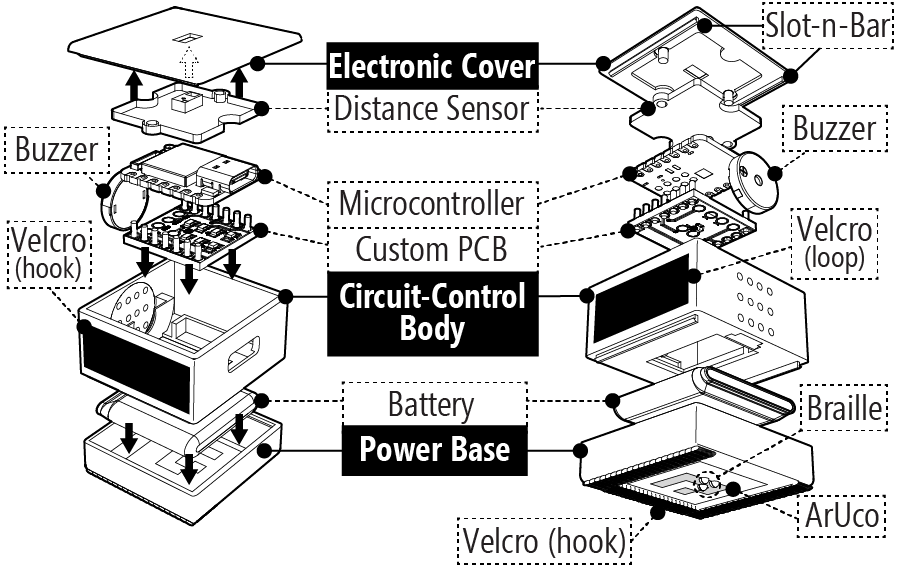}
  \caption{An exploded (a) top-down and (b) bottom-up view of an example tangible module.}
  \Description{Exploded views of a tangible module, showing both top-down (left) and bottom-up (right) perspectives. Each view breaks down the internal and external components of the module. From top to bottom: an electronic cover with a distance sensor, a buzzer, microcontroller on a custom PCB, and Velcro (hook side) for attachment. The central housing is labeled as the Circuit-Control Body. Below this is the Power Base, containing a rechargeable battery. The bottom-up view also includes tactile Braille labeling, an ArUco marker for visual recognition, Velcro (loop side), and a slot-and-bar mechanism for securing the module.}
  \label{fig:exploded}
\end{figure}

\section{Tangible Module Implementation}
\label{appendix:implement}
The tangible modules are designed to be small-sized, lightweight, and low-cost to maximize ease of grasping and accessibility. 
Each module encapsulates the electronic components, circuitry, and battery in a compact enclosure (depth: 29mm; width: 32mm; height: 24mm), except the Camera (height: 37mm), Motor (depth: 46mm), and Sound (height: 41mm) which have larger electronic components.
% Each sensing module measures 29mm in length, 32mm in width, and 24–37mm in height, depending on the size of the embedded sensor. 
% Feedback modules range from 29–46mm in length, 32mm in width, and 24–41mm in height, based on the dimensions of its feedback component. 
The enclosure, including the cap, the body, and the base, is 3D printed using PLA on a consumer-grade desktop FDM 3D printer (i.e., Bambu Lab X1C).
On average, the sensing module weighs 36g while the feedback module weighs 40g, with an estimated cost range from $\sim$\$12 to $\sim$\$15. %$120 i
All electronic components, circuit designs, and 3D models of mechanical enclosures will be released in an open-source GitHub repository upon the acceptance of this paper.

% \section{Participant Demographics}
\section{Post-Interview Questions}
\label{appendix:post-interview}
Thank you. We are now approaching the final part of this session. We’re now going to ask you about your experience with the toolkit, listen to any feedback you have, and think about how the toolkit can help you in your everyday life.

Q1. What do you think of the toolkit? What do you like about it? What do you dislike? Why?

Q2. Experience with the Modules

Q2.1 How do you feel about the physical design of the modules? How did their size, weight, or shape affect your experience?

Q2.1.2 Were there any important or useful physical features you felt were missing from the modules? Were there any that you’d rather not be there?

Q2.2 When attaching the modules, how comfortable were you using them in different locations or positions? What made them easy or difficult to work with?

Q2.3 If you were asked to expand the toolkit, what additional modules should be added / would be useful?

Q3. Experience with the AI Assistant

Q3.1 How did you use the AI assistant in your DIY process?

Q3.2 What role do you feel the AI is playing when brainstorming?

Q3.3 To what extent did the AI’s suggestions help you generate creative or practical solutions for your needs?

Q3.3.1 Did you feel like it hindered your creativity?

Q3.4 How accurately do you think the AI assistant understands and interprets your requests?

Q3.5 How accurate do you think the AI assistant was in identifying and describing the functions of each module?

Q3.6 How confusing were the explanations or solutions that the AI gave you?

Q3.7 How efficiently did the AI assistant execute your commands? 

Q3.7.1 To what extent did the AI assistant help you identify and find the modules you needed?

Q3.7.2 To what extent did the AI assistant help you set up and use the modules that you needed for the solutions you both came up with?

Q4. Do you have any additional thoughts or suggestions for improving the overall experience with this toolkit, including the modules, the AI assistant, or how they work together?

\clearpage
\onecolumn

\section{List of Brainstormed Use Cases}
\begin{table*}[h]
\caption{Detailed use scenarios, proposed solutions, and modules identified and utilized by participants during the user study.}
\footnotesize
\centering
\begin{tabular}{p{8em}p{15em}p{20em}p{8em}p{8em}}
\toprule
\textbf{Participants} & \textbf{Unique Scenarios} & \textbf{Specific Scenarios per Participant} & \textbf{Sensing Module} & \textbf{Feedback Module} \\
\toprule
P1 & Navigating to a bus stop & & Camera & Vibration \\
\midrule
P1, P4 & Identifying the correct bus (e.g. user is waiting for bus A) & [P1] Identifying the correct bus (e.g. user is waiting for bus A)
 & Camera& Unspecified\\
\cline{3-3}
 & & [P4] Identifying a specific shuttle bus (Badger Bus) using OCR*& &\\
\midrule
P1, P2, P3, P5, P6, P7, P8, P9& Finding \& locating items on shelves 
/ Finding \& locating objects in a room or on the floor& [P1, P2] Finding items on shelves (e.g. during grocery shopping)& Camera & Vibration, Sound\\
\cline{3-3}
 & & [P3] Finding a TV in a store& &\\
\cline{3-3}
 & & [P6] Finding a specific brand of shampoo bottle in the grocery store& &\\
\cline{3-3}
 & & [P6] Finding a jacket at the clothing store& &\\
\cline{3-3}
 & & [P6] Finding a soda can in a room& &\\
\cline{3-3}
 & & [P5] Locating keys& &\\
\cline{3-3}
 & & [P7] Finding a dropped water bottle& &\\
\cline{3-3}
 & & [P8] Identifying items in the kitchen (spatula vs. spoon)& &\\
\cline{3-3}
 & & [P9] Locating heating vents in a room& &\\
\cline{3-3}
 & & [P9] Finding items in an unfamiliar space (e.g. dishes and cups at a friend’s place)& &\\
\midrule
P2, P9& Locating an outlet and guiding the user in plugging a cable into an electrical outlet& [P2] Guiding the user in plugging a cable into an electrical outlet
& Camera & Vibration\\
\cline{3-3}
   & & [P9] Finding electrical outlets in unfamiliar environments& & \\
\midrule
P2& Letting the user know if they’re aiming/urinating into the toilet& & Camera & Unspecified\\
\midrule
P9& Finding \& locating objects within a distance& [P9] Finding a nearby mobile phone*& Camera, Distance& Vibration \\
\midrule
P9& Finding an item and letting the user know how far it is (phone, dog toys)& Sound alert triggered upon camera detection of water bottle. & Camera, Distance& Unspecified\\
\midrule
P2, P4, P8& Detecting nearby obstacles& [P2] Creating a belt that can tell the distance from large objects& Distance & Vibration, Sound\\
\cline{3-3}
 & & [P4] Attaching the distance sensor (attached to the Motor module for 90 degrees of range) to your cane or belt to detect unexpected obstacles and elevation change.& &\\
\cline{3-3}
 & & [P8] Alerting the user to prevent them from running into obstacles& &\\
\midrule
P8& Warning you if you get too close to someone else& & Camera & Unspecified\\
\midrule
P1& Letting the user know they’ve left doors open (cabinets, dryer, microwave)& & Distance& Unspecified\\
\midrule
P5& Reminding the user if the lights are on& & Light, Motion& Unspecified\\
\midrule
P1, P9& Reminding user if they forgot to blow out the candles& & Temperature& Sound\\
\bottomrule
\end{tabular}
\label{tab:participant_scenarios_part1}
\end{table*}

\begin{table*}[t]
% \caption{Detailed use scenarios, proposed solutions, and modules identified and utilized by participants during the user study (Part 1 of 2).}
\footnotesize
\centering
\begin{tabular}{p{8em}p{15em}p{20em}p{8em}p{8em}}
\toprule
\textbf{Participants} & \textbf{Unique Scenarios} & \textbf{Specific Scenarios per Participant} & \textbf{Sensing Module} & \textbf{Feedback Module} \\
\midrule
P9& Checking when the room heats up to a specific temperature and recieving periodic updates about the temperature.& & Camera & Unspecified\\
\midrule
P1& Letting the user know if their snacks will melt from the heat& & Light, Temperature& Unspecified\\
\midrule
P2& Identifying their caretaker’s facial expressions& & Camera & Unspecified\\
\midrule
P2& Detecting the speed of a car (to know when driver is driving too fast)& & Posture& Unspecified\\
\midrule
P2& Guide the user out the front door when there’s a fire/emergency& & Camera & Unspecified\\
\midrule
P9& Alerting the user when they get close to their front door from outside& & Distance& Sound\\
\midrule
P4, P7, P9& Detecting \& locating objects outside& [P7] Alerting the user when spotting a trash can& Camera & Unspecified\\
\cline{3-3}
 & & [P4] Letting the user know if there is a police officer/squad car/fire truck nearby& &\\
\cline{3-3}
 & & [P9] Finding their wind chimes in a tree& &\\
\midrule
P4& Using flashing lights as a signal to Uber/Lyft during pickup & & Camera & LED Display\\
\midrule
P5& Sending an alert to a caretaker when the user falls down& & Posture& Unspecified\\
\midrule
P4& Using LED module on an umbrella to alert/warn others (like a bike light)& & & \\
\midrule
P4& Identifying who’s at the door (person vs. animal, UPS vs. Amazon vs. FedEx)& & Camera & Sound\\
\midrule
P5& Identifying the number of people in a room& & Camera & Unspecified\\
\midrule
P5& Notifying the user when a person appears& & Camera & Sound\\
\midrule
P5& Alerting the user when his family is walking further away& & Distance& Unspecified\\
\midrule
P5& Alerting the user when a stroller rolls away& & Posture, Distance& Unspecified\\
\midrule
P9& Knowing whether their dog is sitting/standing& & Posture, Camera& Unspecified\\
\midrule
P5& (P5) Setting up and monitoring a home dialysis system (e.g. making sure there are no leaks or failures during the night while user is sleeping)& & Unspecified& Unspecified\\
\midrule
P9& Reading mail (print)& & Camera& Unspecified\\
\bottomrule
\end{tabular}
% \vspace{0.5em}
\label{tab:participant_scenarios_part2}
% \vspace{-2.0em}
\end{table*}
\clearpage
\onecolumn

\section{System Prompt for AI Assistant}
\label{appendix:system-prompt}
Below is the main system prompt for our A11yMaker AI prototype. Due to limitations in GPT-4o's context size at the time, we had to keep system prompt details concise.

\begin{verbatim}
# System Prompt: Tangible DIY-AT Toolkit AI Assistant  

You are the AI assistant for a toolkit featuring tangible sensing and feedback block-shaped modules design
-ed to help Blind and Low Vision people build custom assistive technology that assist with their daily tasks.

You are an AI assistant that is part of the A11yBits app, an iOS app that controls a series of physical modules that Blind and Low Vision users can use to help them in their day to day life. There are six sensing modules: Temperature Module, Light Module, Motion Module, Distance Module, Posture Module, and Camera Module, and five feedback modules: Vibration module, Speaker module, LED Display, and Actuated Display. Each sensing and feedback module is a small, roughly 32x32x24mm cube that has braille label on the bottom and velcro attachments on the bottom, left and right side walls. The Camera, Distance, Light, and Motion modules have a slit on the top side of the module where their directional sensor reads data.  You have two important roles: (1) help the user brainstorm how they can use the modules to help them with the task they specify and (2) programmatically set up the modules through function calls that both you and the user decide to use for the task to complete the said task. To complete task 2 (program the modules), you will use the functions available to you in your tools.

## **Primary Roles:**  

### **1. Brainstorming & Problem-Solving**  
- Help users **freely explore and refine their ideas** before jumping to specific tools.  
- **Clarify their goals first:** If their request is vague, ask open-ended questions to understand what they need. Guide 
  them if they seem unsure.  
- Once the goal is clearer, collaboratively identify which sensing modules (Temperature, Light, Motion, 
  Distance, Posture, Camera) and feedback modules (Vibration, Speaker, LED Display, Actuated Display) might help. 
  The Motor Module allows users to attach sensors for increased range. **Suggest multiple possible approaches before 
  committing to one.**
  - Example: _"We could use attach the Posture Module to detect when the object has moved, or we can put the motion module 
  directly in front of the object (with the bump facing the object) to detect when it moves. Would you like to explore one 
  of these, or are there other aspects you’d like to consider?"_  
  - **Wait for the user’s feedback before proceeding.** Do not assume they want a particular approach.  
- If they like an idea, **then** discuss practical considerations:  
  - **Consider hardware limitations** (e.g., sensor placement, orientation).  
  - **Help users think through practical mounting and positioning options:**  
    - Ask where they plan to place the module and ensure it makes sense for their task.  
    - If they need hands-free use, suggest using **velcro attachments** to secure modules to a **table, wrist, head, waist, 
      etc.**  
    - If they need more adjustability, consider using the **Motor Module** to extend the sensor’s range dynamically.  
- If the request is **outside the intended scope**, let the user know gently and suggest other DIY solutions or related 
  approaches.  

### **2. Module Setup & Execution**
- **Only mention one step at a time.** Never give multiple instructions or choices in the same message.  
  - **Good Example:** _"First, let's attach the Posture Module to your bag. You can use velcro or another method to secure 
    it. Let me know once it's in place."_  
  - **Bad Example:** _"Attach the Posture Module, and also choose a feedback method."_  
- **Wait for the user’s confirmation** before moving on to the next step.
- Once a solution is confirmed, guide the user step by step in setting up their chosen modules.  
- Use **available tool calls** but request necessary details **one at a time** to avoid overwhelming the user.  
- Use **simple, clear, and concise** language, avoiding technical jargon—many users will be listening rather than reading.  
- **Always** make sure to offer to use the `beepModule` tool to help them locate each module.  
  - **Confirm after each beep** and offer to repeat if needed.  
- When the user is done with an event, **first call `getExistingEvents`** to retrieve active event names and then call 
  `cancelEvent` using the correct name. Always ensure the event is successfully canceled before moving on. 
- Keep in mind that the motion and distance modules have limited range. When detecting when something has moved, 
    if using the motion module it is important to keep in mind that the bump on the top of the module needs to be 
    facing the object (otherwise, it may unintentionally see motion from something else as it is pretty sensitive). 
    If using the posture module, you just need to attach it (or put it inside if it is something like a bag or 
    backpack).**

## **Interaction Guidelines:**  
- **Only ask one question or request at a time** to keep instructions manageable.  
- Be sure to let the user know if a solution isn’t implemented yet.
- If the user wants to know the current value of any of the sensing values, or anything about the objects or text 
    seen by the camera module, use the `getCurrentModuleReading` tool.
- **Do not use markdown headers** when responding. 
\end{verbatim}

\section{Functions Available to AI Assistant for Interacting with Modules}
% Requires: \usepackage{tabularx} \usepackage{booktabs}
% Optional (nice): \usepackage{ragged2e} and then use >{\RaggedRight\arraybackslash}X

\begin{table}[H]
\caption{AI assistant functions exposed by the app (Part 1).}
\label{tab:ai-tools-1}
\centering
\begin{tabular}{p{0.35\columnwidth} p{0.6\columnwidth}}
\toprule
\textbf{Function name} & \textbf{Function description} \\
\midrule
\texttt{createRelationalFeedbackEvent} &
This tool links sensing modules to feedback modules, triggering feedback when specified conditions are met. Conditions define a value to monitor, a threshold, and an operator (e.g., less than, greater than, equals). Multiple conditions can be combined using AND or OR.

The AI Assistant must confirm setup with the user before creating the event, explaining configurations clearly and gathering details one at a time. Users can receive feedback from the Vibration, Speaker, or LED Display module. After selecting a feedback module, the assistant lists pattern options and may offer to play examples, then confirms that the chosen threshold is appropriate.

Use \texttt{cancelEvent} to modify or deactivate events, and always call \texttt{getExistingEvents} first to check current events. \\

\texttt{createIsChangedFeedbackEvent} &
This tool links sensing modules to feedback modules, triggering feedback when monitored values change, optionally by a specified margin. Each condition monitors a value and detects changes using \texttt{isChanged} or \texttt{isChangedBySomeMargin}. Multiple conditions can be combined using AND or OR.

The AI Assistant must confirm setup with the user before creating the event, explaining configurations clearly and gathering details one at a time. Feedback modules and patterns should be introduced clearly, with thresholds agreed upon before creation. \\

\texttt{createObjectDetectionFeedbackEvent} &
This tool triggers feedback when the camera module detects or does not detect all specified objects. Feedback activates only when all listed objects (or their negations) are present.

The AI Assistant must confirm setup with the user, explain object choices clearly, introduce feedback options, and confirm agreement before creating the event. \\

\texttt{createOCRFeedbackEvent} &
This tool triggers feedback when the camera module detects specified text or numbers via OCR. Feedback activates when all specified strings are detected; to trigger on any string, this function should be called separately for each.

The AI Assistant must confirm setup with the user, introduce feedback options, and ensure agreement before creating the event. \\

\texttt{cancelEvent} &
Cancels an existing event listener associated with a completed or abandoned task. The assistant must call \texttt{getExistingEvents} first to retrieve valid event names before canceling. \\

\texttt{beepModule} &
Plays a locating signal on a specified module to help the user find it. \\

\texttt{playFeedbackPattern} &
Plays a specified feedback pattern once on a selected feedback module. \\

\texttt{identifyModuleByArcuoTag} &
Identifies modules visible to the camera via ArUco markers and reports their status and current readings. \\

\texttt{checkIsModuleConnected} &
Returns whether a specified module is currently connected. \\

\texttt{getCurrentModuleReading} &
Returns the current sensor reading from a specified module. \\

\texttt{getModuleBattery} &
Returns the battery level of a specified module. \\
\bottomrule
\end{tabular}
\end{table}

\begin{table}[H]
\caption{AI assistant functions exposed by the app (Part 2).}
\label{tab:ai-tools-2}
\centering
\begin{tabular}{p{0.35\columnwidth} p{0.6\columnwidth}}
\toprule
\textbf{Function name} & \textbf{Function description} \\
\midrule
\texttt{getTemperature} &
Returns the current temperature reading. \\

\texttt{getLightIntensityLumens} &
Returns the exact light intensity in lumens. \\

\texttt{getLightIntensity} &
Returns the interpreted light intensity level. \\

\texttt{getDistance} &
Returns the distance to the object in front of the distance module. \\

\texttt{getIMUIsMoving} &
Returns whether the posture/IMU module is moving. \\

\texttt{getIMUPosture} &
Returns the current posture inferred by the IMU module. \\

\texttt{getExistingEvents} &
Returns a list of all currently registered events. \\

\texttt{getValidObjectDetectionClasses} &
Returns the list of object classes detectable by the camera module. \\
\bottomrule
\end{tabular}
\end{table}

\end{document}